\documentclass[arguments]{aastex631}

\usepackage{hyperref}
\usepackage{amsmath}
\usepackage{enumerate}
\usepackage[utf8]{inputenc}
\usepackage{cleveref}
\usepackage{amssymb}
\usepackage[T1]{fontenc}
\usepackage{enumitem}
\usepackage{url}
\usepackage[misc]{ifsym}
\usepackage{multirow}
\usepackage{mhchem}
\usepackage{booktabs}
\usepackage{CJKutf8}
\usepackage{subfigure}

\begin{document}

\title{Narrowband Radio Technosignature Search toward 3I/ATLAS with FAST}

\begin{CJK*}{UTF8}{gbsn}
\author[0000-0002-1190-473X]{Jian-Kang Li (李健康)}
\affiliation{Institute for Frontiers in Astronomy and Astrophysics, Beijing Normal University, Beijing 102206, People's Republic of China}
 \affiliation{School of Physics and Astronomy, Beijing Normal University, Beijing 100875, People's Republic of China; \url{tjzhang@bnu.edu.cn}}
\author[0000-0002-1190-473X]{Zhen-Zhao Tao (陶振钊)}
 \affiliation{Institute for Astronomical Science, Dezhou University, Dezhou 253023, People's Republic of China}
 \affiliation{College of Computer and Information, Dezhou University, Dezhou 253023, People's Republic of China}
\author[0000-0002-3363-9965]{Tong-Jie Zhang (张同杰)\href{mailto:tjzhang@bnu.edu.cn}{\textrm{\Letter}}}
\affiliation{Institute for Frontiers in Astronomy and Astrophysics, Beijing Normal University, Beijing 102206, People's Republic of China}
\affiliation{School of Physics and Astronomy, Beijing Normal University, Beijing 100875, People's Republic of China; \url{tjzhang@bnu.edu.cn}}



\begin{abstract}
3I/ATLAS is the third confirmed interstellar object passing through the Solar System. In this work, we conduct narrowband radio technosignature search toward 3I/ATLAS using the Five-hundred-meter Aperture Spherical Telescope (FAST) L-band multibeam receiver from October 2025 to January 2026 on 4 separate dates (i.e. Mars closest, perihelion, Earth closest and a post-Earth-closest epoch, respectively). We carry out frequency-drifting signal searching with signal-to-noise ratio (SNR) over 10 within 1.05-1.45 GHz via \texttt{bliss} pipeline. These signal hits are grouped into event by beam, frequency and drift rate matching, the events are then filtered by cluster analysis and drift rate cut-off. We also characterized the events by their significance in SNR, structure tensor as well as principal component analysis (PCA). No credible narrowband radio technosignature are detected from 3I/ATLAS after visual inspections. The null results place constraints on the presence of transmitters above $2.862\times 10^{-3}$ W. We further introduce a Bayesian inference framework to assess the occurrence probability of hypothetical transmitters while accounting for uncertainty in their characteristic transmitter power through physically motivated priors.

\end{abstract}

\keywords{\href{http://astrothesaurus.org/uat/2127}{Search for extraterrestrial intelligence (2127)}; \href{http://astrothesaurus.org/uat/1338}{Radio astronomy (1338)};\href{http://astrothesaurus.org/uat/52}{Interstellar objects(52)}; \href{http://astrothesaurus.org/uat/74}{Astrobiology (74)}}


\section{Introduction}\label{sec:introduction}
\end{CJK*}
The Search for Extraterrestrial Intelligence (SETI) endeavors to identify potential technosignatures as observable manifestations of technology, without presupposing the physical nature of their sources. Narrowband radio emission has long been recognized as a promising technosignature, motivated by the fact that such emissions are not known to arise from natural astrophysical processes \citep{2001ARA&A..39..511T} and can be transmitted efficiently with low energy requirements and minimal propagation losses \citep{1959Natur.184..844C}.

Historically, radio SETI efforts have overwhelmingly focused on stars and planetary systems, particularly those hosting exoplanets \citep{2020AJ....160...29S,2021AJ....161..286T,2021NatAs...5.1148S,2021AJ....161...55M,2022AJ....164..160T,2023AJ....166..190T,2025AJ....169..217L,2026AJ....171...78L}. At the same time, the possibility that interstellar objects or artificial interstellar probes could also constitute relevant technosignature targets proposed by \citet{1960Natur.186..670B} and later discussions by \citet{1985AcAau..12.1027F}. More recently, \citet{2025arXiv250816825D} reviewed the current state of technosignature searches for interstellar objects and related targets. In practice, however, dedicated radio searches of small bodies have remained comparatively rare, largely because such objects are uncommon and typically provide only limited observational windows. The discovery of 3I/ATLAS therefore offers a timely opportunity to extend technosignature searches to this still sparsely explored target class.

3I/ATLAS was discovered on 2025 July 1 by the Asteroid Terrestrial-impact Last Alert System (ATLAS) station. As the third confirmed interstellar object (ISO) traverse the Solar System, similar to the previous confirmed ISO 1I/`Oumuamua and 2I/Borisov, 3I/ATLAS follows an unbound hyperbolic trajectory with eccentricity of 6.1394 and perihelion of 1.3564 AU\footnote{Orbital elements are taken from JPL Horizons: \url{https://ssd.jpl.nasa.gov/horizons/}}. Numerous observations have confirmed its cometary activity, including the presence of a coma \citep{2025ApJ...989L..36S,2025MNRAS.542L.139B,2025RNAAS...9..266F}, a Sun-facing plume \citep{2025MNRAS.544L..31O,2025ApJ...991L..43C,2025A&A...702L...3S,2025ApJ...990L...2J} and a tail \citep{2025ApJ...990L..65K,2025ATel17363....1B}. The reported non-gravitational acceleration also consistent with outgassing-driven forces expected for an active comet\citep{2025RNAAS...9..329E,2025ApJ...990L...2J,2025arXiv251107450N}.

Although the available observational evidence favors a cometary interpretation for 3I/ATLAS, it remains an unusually rare and time-limited target for sensitive radio observations. As the third confirmed interstellar object, it provides a well-defined opportunity to extend narrowband SETI search to a distinct class of nearby transient objects. Several dedicated SETI observations of 3I/ATLAS have already been carried out by Breakthrough Listen utilizing Allen Telescope Array \citep[ATA]{2025arXiv251218142S}, MeerKAT \citep{2025ATel17499....1P} and Green Bank Telescope \citep[GBT]{2025RNAAS...9..351J}, with analyses reported to date yielding null results.

As the largest single-dish radio telescope in the world, FAST can achieve unprecedented sensitivity for detecting faint signal \citep{2019SCPMA..6259502J,2020RAA....20...64J,2020RAA....20...78L,2020Innov...100053Q}. Previous FAST SETI efforts have predominantly targeted extrasolar systems, including nearby stars and globular clusters \citep{2022AJ....164..160T,2023AJ....166..190T,2023AJ....166..245H,2025AJ....169..217L,2026AJ....171...78L,2026AJ....171...51H}. Besides, FAST has also been used to observe Solar System comets, primarily for molecular line studies \citep{2024RAA....24j5008C,2025arXiv251221969J}. 

In this work, we use FAST multibeam L-band observations of 3I/ATLAS obtained at multiple epochs, during which the object's sky position and observing geometry changed, to carry out a dedicated search for narrowband Doppler-drift radio signals, extending narrowband SETI observations to an interstellar comet during its brief passage through the inner Solar System. We introduce Bayes factor to quantify the significance of the difference in SNR between on-source and off-source. Furthermore, we also utilize structure tensor and PCA to characterize the signals we detect for further analysis. In section \ref{sec:Observations}, we discuss the observation setting in this work. The data analysis and signal characterization are introduced in section \ref{sec:DataAnalysis}, and the corresponding results are presented in section \ref{sec:Results}. Section \ref{sec:Discussion} is the implication of our results and section \ref{sec:Conclusion} lists the conclusion of this work.

\section{Observations}\label{sec:Observations}
\subsection{Target Information}
\begin{figure}[htpb]
  \centering
  \includegraphics[width=0.75\textwidth]{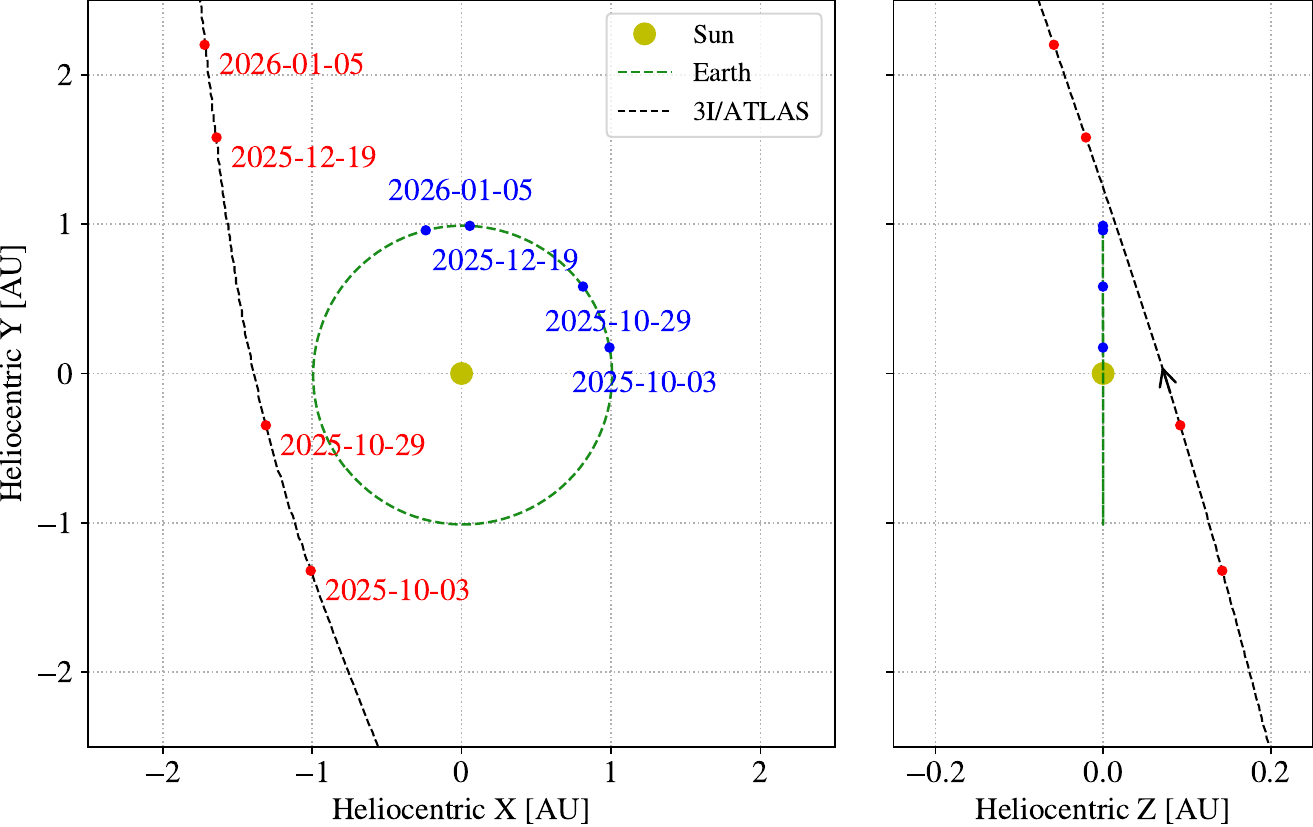}
  \caption{Heliocentric geometry of 3I/ATLAS and the Earth during our observations. The left panel shows the trajectories projected onto the heliocentric ecliptic X-Y plane, while the right panel shows the Z-Y projection.}
  \label{fig:observation_orbit}
\end{figure}

We divide our total observation time into four dates, corresponding to different sky position and observing geometry, and each scan lasts for 1 hour. The details for each observation are listed in Table \ref{table:Target_obs}, and the relative positions of 3I/ATLAS and Earth are illustrated in Figure \ref{fig:observation_orbit}. The four epochs correspond to Mars closest approach, perihelion, Earth closest approach, and a subsequent epoch after Earth closest approach, when the object was receding from Earth.
\begin{deluxetable}{lccccc}[htpb]
\tablecaption{Observation Details of 3I/ATLAS in This Work.\label{table:Target_obs}}
\tablehead{
\colhead{Start Time (UTC)} & 
\colhead{R.A. (J2000.0)} & 
\colhead{Decl. (J2000.0)} & 
\colhead{$\langle d \rangle $ (AU)} & 
\colhead{$\langle \dot{d} \rangle $ (km/s)} & 
\colhead{$\langle \theta_\mathrm{SOT} \rangle (^{\circ})$} 
}
\startdata
2025-10-03 06:00:00 & 14:21:06.43 & -10:37:00.6 & 2.4913 & -6.9589 & 26.7492  \\
2025-10-29 03:30:00 & 13:30:19.35 & -07:00:16.5 & 2.3101 & -17.7519& 12.3551  \\
2025-12-19 21:13:00 & 10:43:02.16 & +07:21:23.5 & 1.7977 & 0.9657 & 108.414  \\
2026-01-05 18:50:00 & 09:22:32.32 & +13:30:16.8 & 1.9446 & 29.7860& 146.2780  
\enddata
\tablecomments{The second and third columns list the mean right ascension and declination for each observation, in the J2000 reference frame. $\langle d \rangle$ and $\langle \dot{d} \rangle$ are the mean geocentric distance between observer and target and its corresponding change rate, respective. $\langle \theta_\mathrm{SOT} \rangle$ is the mean Sun-Observer-Target apparent Solar elongation angle seen from the observer.}
\end{deluxetable}

\subsection{Strategy}\label{subsec:Strategy}

The traditional on-off position-switching strategy is widely used in single-dish radio SETI observations \citep{2017ApJ...849..104E,2020AJ....159...86P,2023AJ....166..206M,2024AJ....167...10C}. In this approach, the telescope alternates between the target position (``on'') and nearby reference pointings (``off''). Artificial technosignatures are expected to be present only in the on-source data and statistically absent in the off-source measurements, which provides a practical discriminator against terrestrial radio frequency interference (RFI) and instrumental artifacts \citep{2017ApJ...849..104E}. A technical description of this strategy as implemented in Breakthrough Listen is given by \citet{2019PASP..131l4505L}.

\begin{figure}[htbp]
    \centering
    \includegraphics[width=0.5\textwidth]{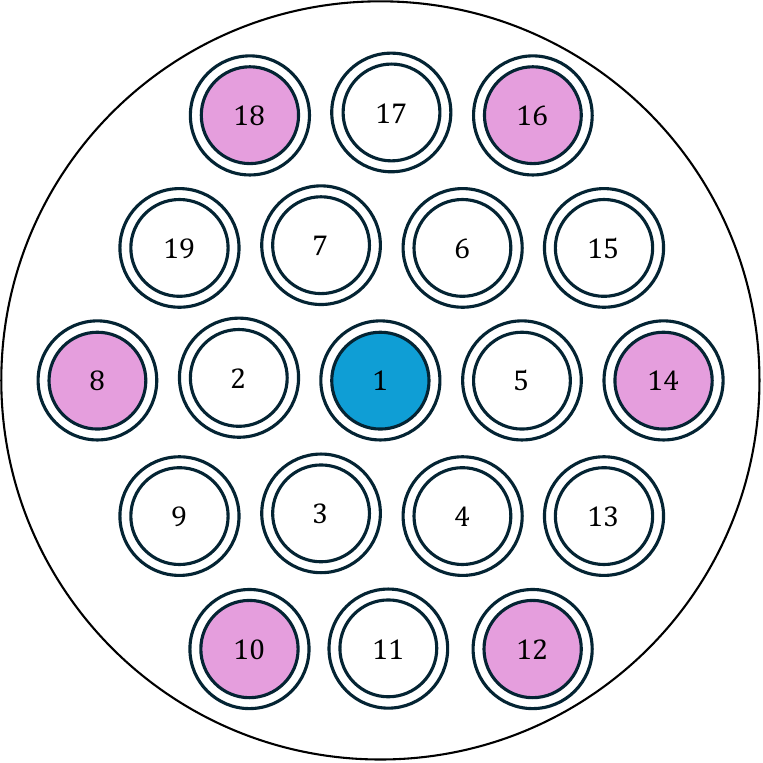}
    \caption{A simple diagram for the multibeam layout of FAST. The central beam (Beam 1) is filled with blue color, and the six outermost beams (Beam 8, 10, 12, 14, 16, 18) filled with pink are the reference beams we use in the search. The angular separation of reference beam is $\sim 11'6$ from the center beam.}
    \label{fig:MBCM}
\end{figure}

The 19-beam receiver of FAST enables a closely related, but more time-efficient, discrimination scheme based on simultaneous multibeam coverage. We adopt a multibeam coincidence matching strategy (MBCM) in which the central beam continuously tracks the target as the on-source beam, while other six outermost beams provide contemporaneous reference measurements (See Figure \ref{fig:MBCM}). Candidate signals are therefore expected to be confined to the on-source beam and absent in the reference beams, allowing effective rejection of broadband or spatially extended RFI through beam-to-beam consistency checks.

\subsubsection{FAST Digital Backend}\label{subsubsec:Backend}
We used the FAST SETI backend, configured with a sampling time of $\sim 10 \mathrm{s}$ and a frequency resolution of $\delta \nu=7.45$ Hz to achieve high spectral resolution. The FAST multibeam digital backend is described in \cite{2020ApJ...891..174Z} and \cite{2020RAA....20...78L}. The data were recorded over 1.0-1.5 GHz with four polarization channels per beam (XX, YY, XY, and YX), corresponding to the auto- and cross-correlations of two orthogonal linear feeds, X and Y. For the analysis we restricted the band to 1.05-1.45 GHz, excluding the outermost 50 MHz at each band edge where the response is degraded by bandpass roll-off. Our narrowband search is performed on all four channels.

\section{Data Analysis}\label{sec:DataAnalysis}

The Doppler-drift search for narrowband signals is performed using \texttt{bliss} \citep{nathan_west_2025_bliss}\footnote{The version of \texttt{bliss} used in this work is 0.0.1rc8.}, a Python package that performs a direct computation of the drift-frequency plane in the drifting signal search. A continuous narrowband signal of extraterrestrial origin is expected to exhibit an apparent frequency drift, $\dot{\nu}\equiv d\nu/dt$, due to relative line-of-sight acceleration between the transmitter and the receiver. We searched drift rates in the range $\dot{\nu}\in[-2,+2]\,\mathrm{Hz\,s^{-1}}$ and SNR threshold of 10.

In our FAST multibeam SETI observations, the central beam continuously tracks the target and is treated as the on-source beam $b_{\rm on}$, while the remaining beams provide simultaneous off-source references. We denote the set of each off-source beams as $\mathcal{B}_{\mathrm{off}}=\{b_{\mathrm{off},i}\}_{i}$, where $i$ is the beam number of the off-source beam (here $i$=8, 10, 12, 14, 16, 18). For a dynamic spectrum $x_b(\nu,t)$, we run a Doppler-drift search over trial drift rates and record a finite set of detections. We define \emph{hit} as any detection returned by the search, characterized by its representative frequency $\nu_0$, trial drift rate $\dot{\nu}_0$, and signal-to-noise ratio $\mathrm{SNR}_h$, satisfying
\begin{equation}
\mathcal{H} \equiv 
\left\{\, h=(\nu_0,\dot{\nu}_0,\mathrm{SNR}_h)\ :\
\mathrm{SNR}_h \ge \mathrm{SNR}_{\rm th}\ \wedge\ 0<|\dot{\nu}_0|\le \dot{\nu}_{\max}
\right\},
\label{eq:hit_def}
\end{equation}
where $\mathrm{SNR}_{\rm th}$ is the signal-to-noise ratio threshold. The hits appear in $b_{\rm on}$ is defined as $\mathcal{H}_\mathrm{on}$, while those appear in $b_{\mathrm{off},i}$ is defined as $\mathcal{H}_{\mathrm{off},i}$. Following the definition in \cite{2021AJ....161..286T}, the \emph{events} are defined as the set of hit presenting in $b_{\rm on}$ over the scan duration $\tau_{\rm obs}$:
\begin{equation}
\mathcal{E}\ \equiv\
\left\{\ 
h\in \mathcal{H}_\mathrm{on}\ : \ |\nu-\nu_0|\le |\dot{\nu}_0|\,\tau_{\rm obs}
\right\}.
\label{eq:event}
\end{equation}
\emph{Potential candidate} is then defined no hits in $\mathcal{B}_{\mathrm{off}}$:
\begin{equation}
\rho\equiv
\left\{
\varepsilon(\nu_0,\dot{\nu}_0)\in\mathcal{E}\ :\
\forall i,\
\nexists\, h'\in \mathcal{H}_{{\rm off},i}\ \text{s.t.}\
|\nu-\nu_0|\le |\dot{\nu}_0|\,\tau_{\rm obs}
\right\}.
  \label{eq:candidate_NB}
\end{equation}

\subsection{Hit Matching, Event Grouping and Potential Candidate Selection}\label{subsubsec:hit_event_cand}
In \cite{2021AJ....161..286T}, on-source and off-source scans are acquired sequentially, so the association between detections is established across scans, and an \emph{event} represents a set of hits in all on-source scans that are consistent in frequency-drift space, and \emph{potential candidate} is defined as event with no hit in all off-source scans \footnote{It should be noticed that the definitions of ``event'' and (potential) ``candidate'' in \citet{2021AJ....161..286T} and this work are relatively unusual than many other SETI literature.}. While in  our multibeam tracking observations, all beams are recorded simultaneously, removing any temporal offset between on and off measurements. In practice, off-source beams may contain no hits within the Doppler-consistent frequency window of an on-source hit, or contain hits that are not consistent in $(\nu_0,\dot{\nu}_0)$ at the search resolution. Strict constraint for candidate that event with no hit in all off-source scans may cause mismatch of hits. 

To avoid this, for each on-source hit $h=(\nu_0,\dot{\nu}_0,\mathrm{SNR}_h)\in\mathcal{H}_{b_{\rm on}}$, we construct a multibeam event by extracting Doppler-consistent dynamic spectrum segments from the on-source beam and from all simultaneously recorded off-source beams, and grouping them into event by rewritten Eq. (\ref{eq:event}) as
\begin{equation}
\mathcal{E}_{\rm MB}\ \equiv\
\left\{ \varepsilon_{\rm MB}(h,x_b)=\begin{pmatrix}
h(\nu_0,\dot{\nu}_0) \\
\left\{x_b(\nu,t)\right\}_{b\in \mathcal{B}_{\rm off}}
\end{pmatrix}\  :
h\in \mathcal{H}_\mathrm{on}\ \wedge \ |\nu-\nu_0|\le |\dot{\nu}_0|\,\tau_{\rm obs}
\right\}.
\label{eq:event_def}
\end{equation}
For a given on-source hit $h=(\nu_0,\dot{\nu}_0)\in\mathcal{H}_{\rm on}$ and the $i$th off-source beam with hit set $\mathcal{H}_{{\rm off},i}$, we define a matched \emph{detection} if there exists $h'=(\nu'_{0},\dot{\nu}'_{0})\in\mathcal{H}_{{\rm off},i}$ such that
\begin{equation}
  \mathcal{D}_i(\nu_0,\dot\nu_0)\ \equiv\
\left\{
x_b(\nu,t)\ :\
\forall b\in \mathcal{B}_{\rm off}, \ \exists\, h'(\nu'_0,\dot{\nu}'_0)\in \mathcal{H}_{{\rm off},i}\ \ \text{s.t.}\ \
\left|\nu'_0-\nu_{0}\right|\le \delta \nu\ \wedge\
\left|\dot{\nu}'_0-\dot{\nu}_{0}\right|\le \delta \dot{\nu}.\right\}
\label{eq:detection_def}
\end{equation}
where $\delta \dot{\nu}=\delta \nu/\tau_{\rm obs}$ is the drift rate resolution ($\delta \dot{\nu}=0.0024$ Hz/s in this search). And the potential candidate is therefore defined by 
\begin{equation}
  \rho_{\rm MB}\equiv
\left\{
\varepsilon_{\rm MB}(h,x_b)\in\mathcal{E}_{\rm MB}:\
\forall\, b\in \mathcal{B}_{\rm off} ,\ x_b \notin \mathcal{D}_i(\nu_0,\dot{\nu}_0)
\right\}.
\label{eq:candidate_def}
\end{equation}

\subsection{Hits Filtering}\label{subsubsec:event_filter}

The hits we get after hit searching are not the final hits used for event grouping. Because RFI phenomenology is strongly frequency dependent, we perform clustering independently within coarse frequency intervals. We partition the band into non-overlapping bins of width 50 MHz. Within each bin, we combine on-source hits and the available off beams. To prevent the clustering structure from being dominated by the typically much larger off-source population, we perform a balanced subsampling of the off-source hits within each 50 MHz frequency bin. Specifically, in each bin we keep all on-source hits and select an off-source subset with a total size comparable to the on-source sample, while preserving the relative representation of the different off beams. We then apply Hierarchical Density-Based Spatial Clustering of Applications with Noise \citep[HDBSCAN;][]{10.1007/978-3-642-37456-2_14} within each bin to the feature vectors $(\nu,\dot{\nu},\log_{10}{\rm SNR})$ after robust scaling. Similar density-based clustering strategies have recently been explored in SETI works. For example, \citet{2025AJ....169..206J} applied HDBSCAN to features including frequency, drift rate, and $\log(\mathrm{SNR})$ to group morphologically similar hits, while \citet{2020ApJ...891..174Z,2026AJ....171...86Z} used DBSCAN for residual RFI mitigation in SETI commensal survey data. Besides recording the cluster label for each hit, we also record the membership probability that quantifies how strongly a point belongs to its assigned cluster. For each cluster, we additionally compute summary diagnostics such as the cluster size and the on/off composition, including the off fraction
\begin{equation}
  f_{\rm off}(c)=\frac{N_{\rm off}(c)}{N_{\rm on}(c)+N_{\rm off}(c)},
\end{equation}
where $N_{\rm on}(c)$ and $N_{\rm off}(c)$ denote the numbers of on-source and off-source hits assigned to cluster $c$, respectively. The on-source hits that fall into clusters with large $f_{\rm off}$ are interpreted as morphologically consistent with the off-source population and thus more likely to be RFI-like, whereas on-source hits assigned to clusters with small $f_{\rm off}$ or labeled as noise are treated as atypical and prioritized for further vetting (here we set the $f_{\rm off}$ threshold of 0.1\%).

We also filter the hits by drift rate before potential candidate selection. The drift rates caused by the relative motion of comet in the observation dates are listed in Table \ref{table:Target_motion}. Although the maximum drift rate (MDR) we used is 2 $\mathrm{Hz \, s^{-1}}$ in the hit searching, the relative accelerations of the comet are at about $\sim \, 10^{-5}\ km\,s^{-1}$ level during our observation within our observation frequency range, and correspond to drift rate of approximately 0.1-0.2 $\mathrm{Hz \, s^{-1}}$. The rotation period of 3I/ATLAS is 16.16 hours \citep{2025A&A...702L...3S}, and the nucleus radius is estimated as $\sim$ 3 km \citep{2025ApJ...990L..65K,2025A&A...702L...3S}, resulting in a rotational acceleration of $\sim 3.50\times10^{-5}\ \mathrm{m\,s^{-2}}$ and corresponding drift rate of $\sim 1.23-1.69\times10^{-4} \ \mathrm{Hz\,s^{-1}}$ within our observation frequency range, which is 2 magnitudes of order higher than the relative accelerations in Table \ref{table:Target_motion}, and therefore can be neglected.
\begin{deluxetable}{lccc}[htpb]
\tablecaption{Accelerations and drift rates of 3I/ATLAS during our observation dates.\label{table:Target_motion}}
\tablehead{
\colhead{Observation Date} & 
\colhead{$\langle \dot{v} \rangle (\mathrm{km/s^2})$ } & 
\colhead{$\langle \dot{\nu} \rangle $ at 1250 MHz [$\mathrm{Hz \, s^{-1}}$]} & 
\colhead{Used $\langle \dot{\nu} \rangle $ range [$\mathrm{Hz \, s^{-1}}$]} 
}
\startdata
2025-10-03 & $2.632\times10^{-5}$ & 0.109 & (0,0.15) \\
2025-10-29 & $2.553\times10^{-5}$ & 0.105 & (0,0.15)   \\
2025-12-19 & $4.794\times10^{-5}$ & 0.199 & (0,0.25)   \\
2026-01-05 & $5.002\times10^{-5}$ & 0.207 & (0,0.25)   
\enddata
\tablecomments{The second and third columns are the mean acceleration and drift rate at central frequency for each observation, the fourth column is the drift rate we used for filtering.}
\end{deluxetable}

\subsection{Significance of SNR}\label{subsubsec:SNR_significance}
The modified definitions for event also enable us to quantify the significance in SNR between on-source and off-source hits. In multibeam tracking observations, signals appear in multiple beams within a certain frequency range are most probably RFIs. Therefore, beyond the binary criterion of the existence of counterpart off-source hits, we can define a quantitative metric that measures how consistent the strength of the on-source and off-source detections. In particular, a candidate-like signal is expected to be significantly stronger in the on-source beam than in off-source beam, whereas an RFI-like signal tends to show comparable strengths across beams. 

During the event grouping, we also record the SNR of the hits in the event. For each off-source dynamic spectrum in the set $\left\{x_b(\nu,t)\right\}_{b\in \mathcal{B}_{\rm off}}$, if there are more than one hits satisfy Eq. (\ref{eq:detection_def}), we select the hit with the SNR closest to the SNR of corresponding on-source hit among these off-source hits, and its detection status is marked as ``1'' (All on-source hits are mark as ``1''). While if the off-source dynamic spectrum satisfy $x_b \notin \mathcal{D}_i(\nu_0,\dot{\nu}_0)$ within the frequency range, its detection status is marked as ``0'' and we select the SNR threshold as an upper limit. So potential candidate should be those events with detection status of 1000000. Since there are small gain differences between the central beam and the reference beams, we apply a beam gain ratio adjustment to the off-source SNR values and SNR threshold upper limits used in the event-level SNR significance calculation by the beam gain ratio $R_b(\nu)\equiv G_b(\nu)/G_{\rm on}(\nu)$, which can be obtained from the weekly updated antenna gain measurement of FAST. The off-center beams have relative beam efficiencies mostly above 90\% of the central beam, and the polarized beam responses of the off-center beams do not differ from the central beam at an order-of-magnitude level \cite{2025AJ....169..158C}. The polarization measurement for the 19 beams by \citet{2025AJ....170..116C} reveals that the central-beam polarization calibration is more accurate than that of the off-center beams, however, the resulting Mueller-matrix parameters and the terms most relevant to the cross products, such as the residual phase, feed ellipticity, and leakage, remain small corrections rather than large changes in response. We therefore assume, in this rough SNR reference adjustment for off-source beams, that the beam-to-beam cross-correlation products XY/YX response ratio is of the same order as the auto-correlation XX/YY or Stokes-$I$ gain ratio $R_b(\nu)$\footnote{Gain measurements of FAST: \url{https://fast.bao.ac.cn/cms/category/1977569564215562242}. The ratios are used only as an approximate reference adjustment for comparing off-source SNRs with the central-beam SNR, and do not affect the event selection. The gain corrections for cross-correlation products need more accurate and detailed polarization measurements and calibrations.}.

Based on statistical analysis of past multibeam observations with SNR threshold of 10 \citep{2022AJ....164..160T,2023AJ....166..190T,2025AJ....169..217L}, we can determine that the distribution of SNRs can be modeled by a log-normal distribution $\ln y \sim \mathcal{N}(\mu,\sigma^2)$. We then compare two competing hypotheses for each event: (i) $H_{\rm R}$ (RFI-like): the SNR values of on-source and off-source are drawn from the same parent distribution; (ii) $H_{\rm O}$ (on-source-enhanced): the on-source SNR value is systematically stronger than the off-source ensemble by an offset $\Delta \ge 0$. These two hypotheses can be formulated by 
\begin{equation}
  H_{\rm R}:\quad \ln y_{\rm on} \sim \mathcal{N}(\mu,\sigma^2),\quad
\ln y_{b} \sim \mathcal{N}(\mu,\sigma^2),
\label{eq:H_sFI}
\end{equation}
and
\begin{equation}
  H_{\rm O}:\quad \ln y_{\rm on} \sim \mathcal{N}(\mu+\Delta,\sigma^2),\quad
\ln y_{b} \sim \mathcal{N}(\mu,\sigma^2),\quad \Delta\ge 0.
\label{eq:H_cand}
\end{equation}
The likelihood for $H_{\rm R}$ can be written by 
\begin{equation}
  \mathcal{L}(D \mid \mu,\sigma, H_{\rm R}) =
p(y_{\rm on}\mid \mu,\sigma)\,
\prod_{b\in \mathcal{B}_{\rm det}} p(y_b\mid \mu,\sigma)\,
\prod_{b\in \mathcal{B}_{\rm nodet}} P(y_b \le L_b\mid \mu,\sigma),
\end{equation}
and the likelihood for $H_{\rm O}$ can be written by
\begin{equation}
\mathcal{L}(D \mid \mu,\sigma,\Delta, H_{\rm O}) =
p(y_{\rm on}\mid \mu+\Delta,\sigma)\,
\prod_{b\in \mathcal{B}_{\rm det}} p(y_b\mid \mu,\sigma)\,
\prod_{b\in \mathcal{B}_{\rm nodet}} P(y_b \le L_b\mid \mu,\sigma),
\end{equation}
where $\mathcal{B}_{\rm det}$ are the beams with detection status of ``1'', while $\mathcal{B}_{\rm nodet}$ are the beams with detection status of ``0''. We quantify the strength-contrast evidence via the Bayes factor
\begin{equation}
  \mathrm{BF} \equiv \frac{p(D\mid H_{\rm O})}{p(D\mid H_{\rm R})},
\label{eq:BF_def}
\end{equation}
where the marginal likelihood under each hypothesis is obtained by integrating over the nuisance parameters,
\begin{equation}
  p(D\mid H_{\rm R}) = \int p(D\mid \mu,\sigma, H_{\rm R})\,p(\mu,\sigma)\,d\mu\,d\sigma,
\end{equation}
and 
\begin{equation}
  p(D\mid H_{\rm O}) = \int p(D\mid \mu,\sigma,\Delta, H_{\rm O})\,p(\mu,\sigma,\Delta)\,d\mu\,d\sigma\,d\Delta.
\end{equation}
For the hyperparameters $\mu$, $\sigma$ and $\Delta$, we adopt weakly informative hyperpriors
\begin{equation}
  \mu \sim \mathcal{N}(\mu_\mu,s_\mu^2), \quad
\sigma \sim \mathrm{HalfNormal}(s_\sigma), \quad
\Delta \sim \mathrm{HalfNormal}(s_\Delta),
\end{equation}
with $(\mu_\mu,s_\mu,s_\sigma)$ estimated from the off-source beams hits in the same observation.

\subsection{Structure Tensor}\label{subsubsec:structure_tensor}
The waterfall plot of a dynamic spectrum may be viewed as a 2-dimension scalar field image $I(\nu,t)$, where each pixel corresponds to $\mathrm{p}=(\nu_i,t_i)$. In the presence of noise, a narrowband signal with a nonzero drift rate introduces local structure with both directionally organized and anisotropic feature within a small neighborhood in the $(\nu,t)$ plane. Such local orientation can be quantified by the gradient structure tensor \citep{2011LNCS.6688..545K}, i.e., the second-moment matrix of the image gradients,
\begin{equation}
J(\mathrm{p})\equiv
\left\langle \nabla I(\mathrm{p})\,\nabla I(\mathrm{p})^T\right\rangle
=
\begin{pmatrix}
J_{tt} & J_{t\nu}\\
J_{t\nu} & J_{\nu\nu}
\end{pmatrix}
=
\begin{pmatrix}
\langle I_t^2\rangle & \langle I_t I_\nu\rangle\\
\langle I_t I_\nu\rangle & \langle I_\nu^2\rangle
\end{pmatrix},
\end{equation}
where $\nabla I=(I_t, I_\nu)^T$ with $I_t\equiv \partial I/\partial t$ and $I_\nu\equiv \partial I/\partial \nu$, and $\langle\cdot\rangle$ denotes a local neighborhood average. The eigenvalues of $J$ are
\begin{equation}
\lambda_{1,2}=\frac{1}{2}\left[(J_{tt}+J_{\nu\nu})\pm
\sqrt{(J_{tt}-J_{\nu\nu})^2+4J_{t\nu}^2}\right].
\end{equation}
A convenient scalar measure of local anisotropy is the coherence,
\begin{equation}
coh=\frac{\lambda_1-\lambda_2}{\lambda_1+\lambda_2},
\end{equation}
which ranges from $0$ (locally isotropic texture, with no preferred direction) to $1$ (a local structure whose gradients are strongly aligned with a dominant direction).

The gradient orientation angle can be written as
\begin{equation}
\theta_g=\frac{1}{2}\arctan\frac{2J_{t\nu}}{J_{tt}-J_{\nu\nu}},
\end{equation}
and the orientation of the underlying line-like feature (i.e., the local ridge direction) is orthogonal to the gradient, such that
\begin{equation}
\theta_{\rm line}=\theta_g+\frac{\pi}{2}.
\end{equation}
With $\theta_{\rm line}$ defined relative to the frequency axis in the $(\nu,t)$ plane, the corresponding local slope yields an estimate of the drift rate,
\begin{equation}
\dot{\nu}=\frac{d\nu}{dt}=\tan\theta_{\rm line}\,\frac{\delta\nu}{\delta t},
\end{equation}
where $\delta\nu/\delta t$ converts pixel units to physical units. Intuitively, for a small dynamic spectrum patch containing a line- or ridge-like feature, the local angle describes the dominant orientation of that feature (and hence its local drift direction), whereas the coherence measures how strongly the patch resembles a well-defined line rather than an isotropic texture.

\begin{figure}[htpb]
  \centering
  \includegraphics[width=0.9\textwidth]{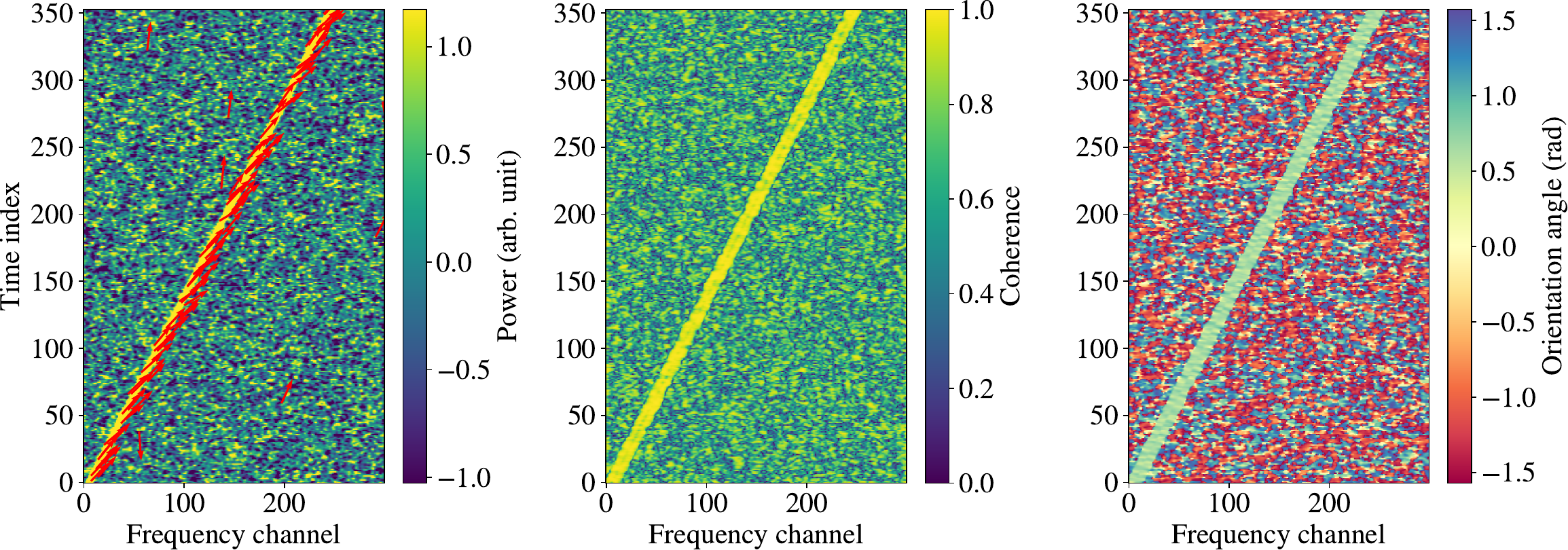}
  \caption{\label{fig:Structure_tensor_onscan}  Waterfall plots for a dynamic spectrum with a frequency drifting signal (left), its corresponding coherence map (middle) and orientation angle (right). Quivers are plotted in the high-coherence region (${\rm coh}>0.97$). The contiguous, dense quiver field traces the signal track, and the arrow orientation encodes the sign of the drift rate.}
\end{figure}

In practice, to improve robustness against noise, we apply two stages of Gaussian smoothing. We first smooth the dynamic spectrum with widths $(\sigma_{\rm pre,t},\sigma_{\rm pre,\nu})$ before computing numerical gradients, and then compute the tensor components from locally averaged outer products of the gradients using a second smoothing with widths $(\sigma_{\rm st,t},\sigma_{\rm st,\nu})$. Optionally, a non-negative weight map $w(p)$ may be introduced in the outer products (e.g., based on local signal energy) to emphasize high-power regions when estimating $J$. Figure~\ref{fig:Structure_tensor_onscan} illustrates the structure tensor characterization of a simulated drifting narrowband signal generated with \texttt{setigen} \citep{2022AJ....163..222B} and embedded in Gaussian noise. The injected signal has a Gaussian spectral profile with width $3\delta\nu$, an amplitude corresponding to $\mathrm{SNR}=40$, and a drift rate of $+0.5\,\mathrm{Hz\,s^{-1}}$.

For this simulated example, the high-coherence region gives an estimated drift rate of $\dot{\nu}\sim0.46\,\mathrm{Hz\,s^{-1}}$ by $\dot{\nu} =(\delta\nu/\delta t)\tan\theta$, generally consistent with the injected value of $+0.5\,\mathrm{Hz\,s^{-1}}$. The small difference is expected because the estimate is obtained from local gradients in a finite, noisy, discretely sampled dynamic spectrum rather than from the known injected track itself. In addition, the long-duration simulation makes the drifting line span numerous frequency channels, so edge truncation, finite pixel sampling, local smoothing, and accumulated noise fluctuations can slightly bias the orientation-based drift estimate.

\subsection{Principal Component Analysis}\label{subsubsec:PCA}
The dynamic spectrum can be also regarded as a 2-dimension matrix $X \in \mathbb{R}^{N_t \times N_\nu}$, then we can extract the principal components (PCs) for this dynamic spectrum, and featurize the drifting signals. PCA is an unsupervised machine-learning method for data decomposition or compression. For our 2-dimension dynamic spectrum matrix, we can use singular value decomposition (SVD) to decompose the matrix into
\begin{equation}
  X_c = U\,\Sigma\, V^T,
\end{equation}
where $X_c$ is the matrix after mean bandpass filtering,
\begin{equation}
\mu_j = \frac{1}{N_t}\sum_{i=1}^{N_t} X_{ij},\quad
X_c = X - \mathbf{1}\mu^T,
\end{equation}
$U\in\mathbb{R}^{N_t\times r}$ and $V\in\mathbb{R}^{N_\nu\times r}$ are the unitary matrix, $\Sigma=\mathrm{diag}(\sigma_1,\ldots,\sigma_s)$ is the rectangular diagonal matrix with $\sigma_i$ being the singular values of $X_c$, and $r\le \min(N_t,N_\nu)$ is ranks. Such mean bandpass filtering can efficiently remove the time-invariant (non-drifting) components at each frequency channel. Consequently, stationary carriers and persistent bandpass-like features are strongly suppressed, while frequency-drifting tracks remain in $X_c$. We can then reconstruct the dynamic spectrum by 
\begin{equation}
  \widehat{X_{k}} = U_k\,\Sigma_k\, V_k^T+\mathbf{1}\mu^T,
\end{equation}
with explained-variance ratio $\mathrm{EVR}_k=\sigma_k^2/\sum_j\sigma_j^2$. The $k$-th PCs are given by the columns of $V$ ($v_k\in\mathbb{R}^{N_\nu}$), which describes the frequency-domain loading pattern. The corresponding temporal score is $S=U\Sigma$ ($s_k=\sigma_k u_k \in \mathbb{R}^{N_t}$), where $u_k$ is the $k$-th column of $U$. Thus, the temporal score gives the time-dependent amplitude with which the frequency-domain component $v_k$ contributes to the dynamic spectrum. For a Doppler-drifting narrowband line in a dynamic spectrum, the EVR of the PCA can be approximated by

\begin{equation}
\mathrm{EVR}_m \propto \exp\left[-\left(\frac{2\pi m w_\nu}{|\dot\nu|\tau_{\rm obs}}\right)^2\right], \quad m=0,1,\dots ,
\label{eq:EVR}
\end{equation}
where $w_{\nu}\ll |\dot\nu|\tau_{\rm obs}$ denotes the effective linewidth in frequency of the drifting feature, and $m$ is the Fourier-mode order. We further define the EVR entropy as
\begin{equation}
  H_{\rm EVR}=-\sum_m p_m\ln p_m,\quad p_m=\frac{\mathrm{EVR}_m}{\sum_j \mathrm{EVR}_j}.
\end{equation}
Under the above approximation, the EVR entropy of a drifting line can be expressed as
\begin{equation}
H_{\rm EVR}\approx\ln\left(\frac{|\dot\nu|\tau_{\rm obs}}{2\pi w_\nu}\right)+\frac{1}{2}\ln(\pi e).
\label{eq:H_EVR}
\end{equation}
The detailed derivations of the PCA for Doppler-drifting signal are in Appendix \ref{subsec:appendix_EVR_PCA}. Figure \ref{fig:PCA_onscan_example} is the PCA summary for a simulated drifting signal used in Section \ref{subsubsec:structure_tensor}. 
\begin{figure}[htpb]
  \centering
  \includegraphics[width=0.9\textwidth]{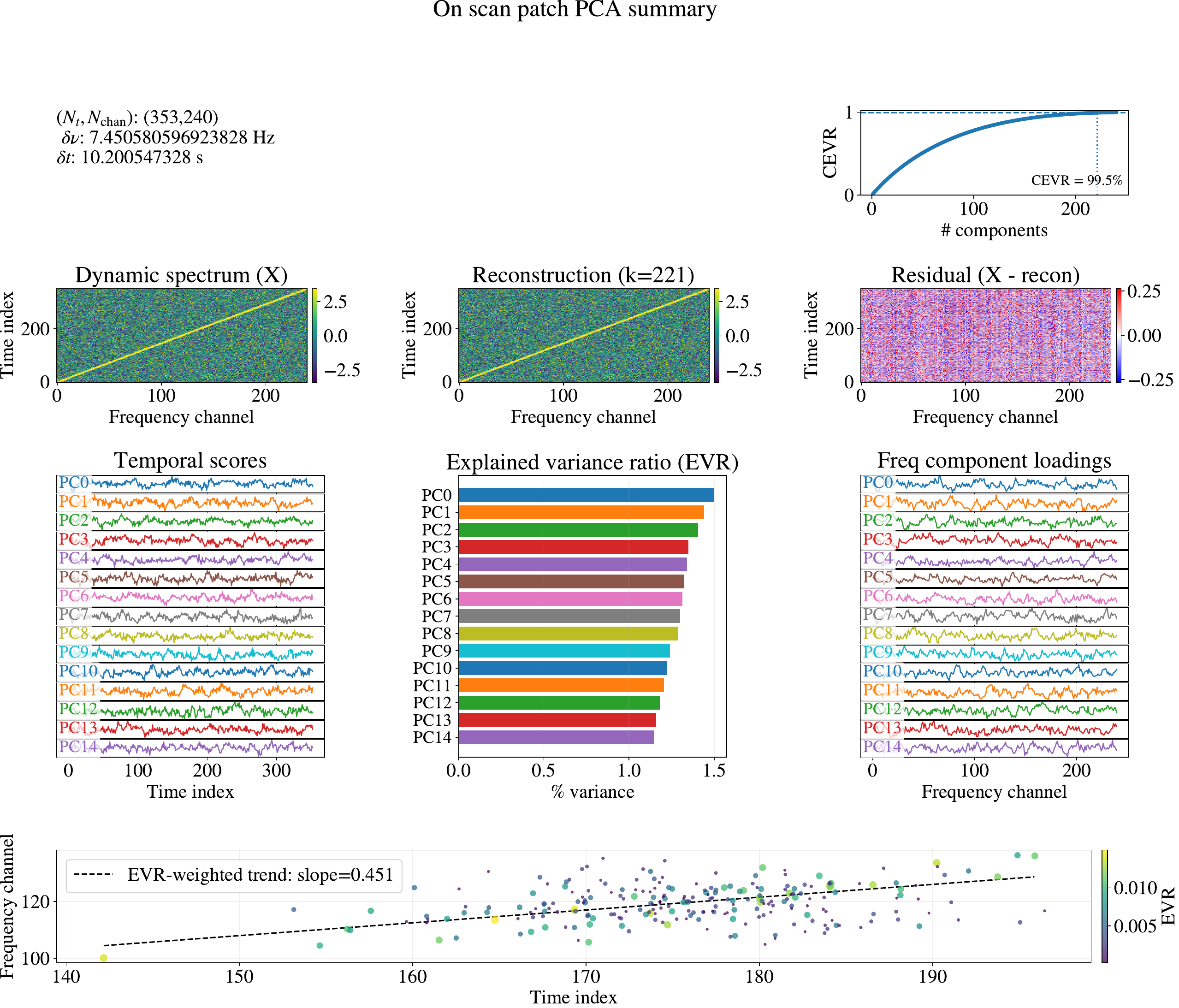}
  \caption{\label{fig:PCA_onscan_example}  PCA summary for the drifting signal. First row: metadata for this panel (left) and the cumulative explained variance ratio (CEVR) curve (right). We define the CEVR as $\sum_{k=1}^{K}\mathrm{EVR}_k$. The vertical dotted line marks the minimum number of PCs required to reach $\mathrm{CEVR}\ge 0.995$. Second row: the original dynamic spectrum (left), its rank-$k_{\rm recon}$ reconstruction $\hat X_{k_{\rm recon}}$ (center), and the residual $X-\hat X_{k_{\rm recon}}$ (right). Third row: PCA diagnostics for the first $k_{\rm recon}$ temporal score loadings $s_k(t)$ (left), explained variance ratio $\mathrm{EVR}_k$ (center), and frequency component loadings $v_k(\nu)$ (right). Scores and components are normalized for display, the color scheme is consistent across the three bottom panels for each PC. Fourth row: time-frequency loading diagnostic illustrating whether PCs are arranged along a drift-like time-frequency trend. Each point represents the power-weighted response centroid of one principal component in the \((t,\nu)\) plane, and the dashed line is a linear fit to these points with the EVR used as the fitting weight. Although the top component number for reconstruction is 221, we only show the top 15 components in the third row.}
\end{figure}
For a drifting narrowband signal, a certain number of PCs typically captures a correlated spectral pattern $v_k(\nu)$ and a smoothly varying temporal score $S_k(t)$, producing a compact representation in EVR. The EVR entropy for the simulated signal is 5.094, which is modestly larger than the theoretical approximation values 4.722 in Eq. (\ref{eq:H_EVR}) with $w_\nu\approx3\delta\nu$. This difference is expected because the simulated data do not exactly satisfy the ideal shift-invariant Gaussian-kernel assumption. Finite drift span and edge truncation can introduce discrete spectral leakage, small time variability in the line amplitude or profile can spread power across additional modes, and residual correlated noise or spectral texture can further transfer variance into higher-order PCs. Together, these effects raise the measured $H_{\rm EVR}$ modestly above the theoretical estimate. 

To examine whether the PCA decomposition captures a possible frequency-drifting pattern, we assign each PC a representative point in the time--frequency plane. For the $k$-th component, with temporal score $s_k(t)$ and frequency loading $v_k(\nu)$, we define
\begin{equation}
t_k =
\frac{\sum_t t |s_k(t)|^2}
{\sum_t |s_k(t)|^2},
\qquad
\nu_k =
\frac{\sum_\nu \nu |v_k(\nu)|^2}
{\sum_\nu |v_k(\nu)|^2}.
\end{equation}
The set of points $(t_k,\nu_k)$ provides a compact time--frequency loading diagnostic of the PCA decomposition. Since $t_k$ and $\nu_k$ are power-weighted response centroids rather than local track coordinates, these points naturally tend to cluster near the time--frequency region where the integrated PC response is strongest. The two ends of a drifting track can still contribute through the global response weights of the corresponding scores and loadings, but they are not represented as separate local trajectory points. We then fit a linear relation $\nu_k=a_{EVR} t_k+b_{EVR}$ to these points, using the EVR of each component as the fitting weight. This diagnostic is used only to qualitatively examine whether the representative PC response locations show an overall frequency-drifting tendency, rather than to reconstruct the full drifting track.

\subsection{Overall Workflow}\label{subsec:workflow}
\begin{figure}[htpb]
  \centering
  \includegraphics[width=0.85\textwidth]{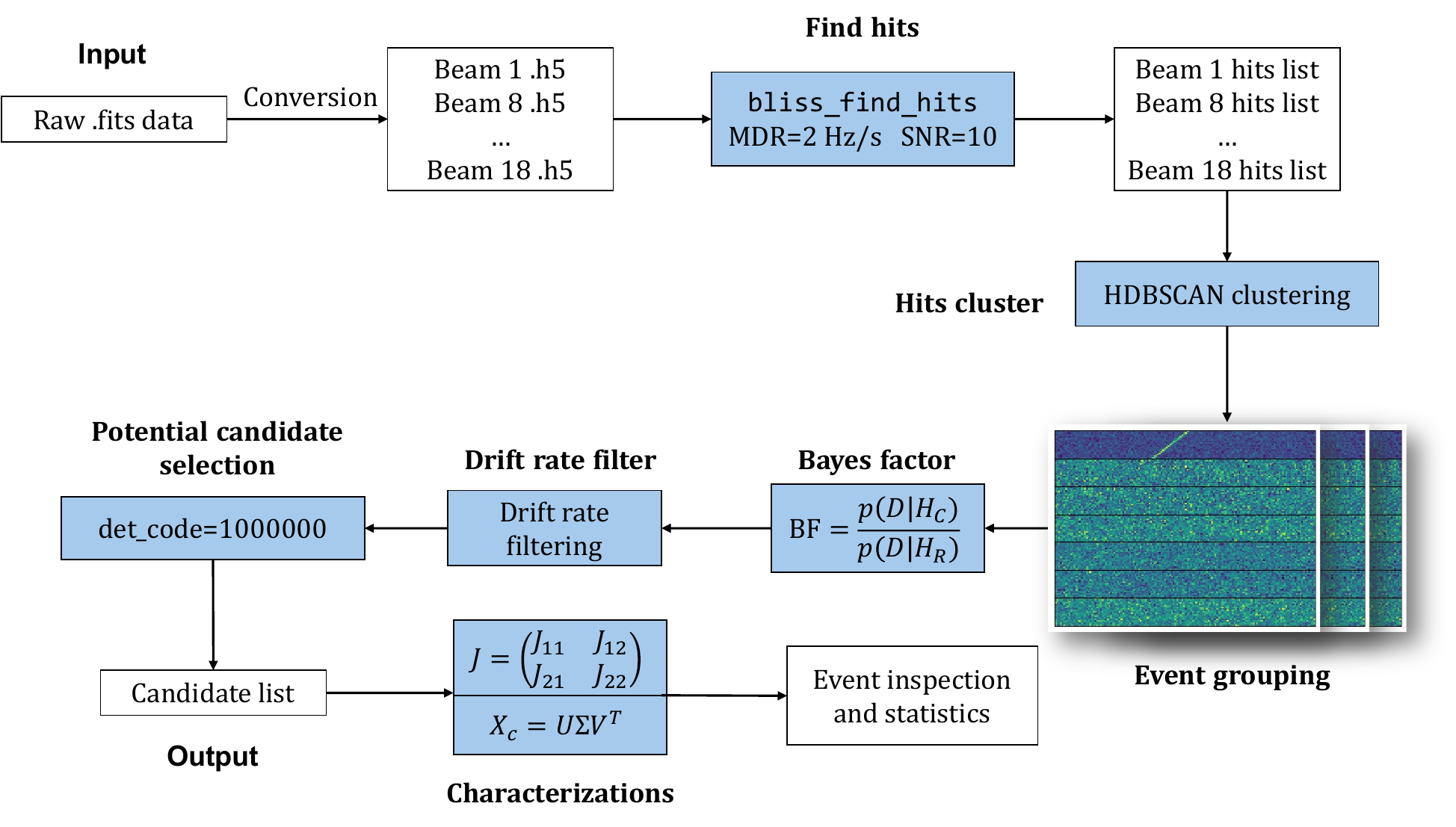}
  \caption{\label{fig:overall_pipeline}  A schematic representation of our workflow for the data analysis.}
\end{figure}

The overall data analysis processes are illustrated in Figure \ref{fig:overall_pipeline}. Before the finding hits step with \texttt{bliss\_find\_hits}, the raw sdfits data are converted into HDF5 format by \texttt{blimpy} \citep{2019JOSS....4.1554P}. The \texttt{fits} files are archived for each beam, each \texttt{fits} file corresponds to a different sample time segment of the observation. For each beam, these \texttt{fits} files are time-ordered, concatenated along the time axis, and then converted into an HDF5 product containing two-dimensional time-frequency power spectra for four polarization channels, XX, YY, XY and YX, which are derived from self-correlation and cross correlation of the data in two orthogonal linear polarization directions X and Y. 

In this work, we perform an independent narrowband Doppler-drift search on each of the four polarization channels and the Stokes-$I$ total intensity, following the procedure outlined in Figure \ref{fig:overall_pipeline}. For each channel, we generate a list of potential candidates and inspect them using standard dynamic spectrum (waterfall plot) visualizations to assess whether they meet the criteria in Eq. (\ref{eq:candidate_def}). This assessment is complemented by structure-tensor diagnostics and PCA summary plots, which provide additional feature-based characterization of the signals. We then quantify the physical and statistical properties of the resulting sample using dedicated statistical analyses.

\section{Results}\label{sec:Results}
The general number statistics for hits, events and potential candidates in this narrowband drifting signal search are listed in Table \ref{table:search_stat}. There are two major types of RFI sources at the FAST site, civil aviations and navigation satellites, which occupy 22.5\% and 14.76\% of the total bandwidth (See \cite{2021RAA....21...18W} for detailed frequency coverage). About 40\% hits fall in the frequency ranges of civil aviation and navigation satellites, implying that a considerable amount of RFIs may come from these known sources. The larger hit fraction in the navigation satellite bands, compared with their bandwidth fraction, is expected because these transmitters are usually strong and persistent RFI sources, which is also consistent with the RFI environment tests at the FAST site.

We further decompose the event-level filtering stage in Table~\ref{table:search_stat}. A potential candidate is required to have no matched off-hit counterpart, not to be assigned to large off-hits populated cluster, and to fall within the adopted date-dependent drift-rate range simultaneously. The auto-correlation products XX and YY have filtering fractions close to those of Stokes-$I$, whereas the cross-correlation products, especially XY, show different behavior. In particular, the larger no-off-hit and unclustered fractions in XY leads its relatively high candidate fraction.

\begin{deluxetable}{lccccccc}[htpb]
\tablecaption{Numbers of hits, events, and potential candidates in the narrowband drifting signal search.\label{table:search_stat}}
\tablehead{
\colhead{Polarization} & 
\colhead{Hits} &
\colhead{Hits in Civil, GNSS (\%)} &
\colhead{Event} &
\colhead{No off hit (\%)} &
\colhead{Unclustered (\%)} &
\colhead{Drift range (\%)} &
\colhead{Candidate}
}
\startdata
Stokes-$I$ & 7743123 & 2.46, 33.94 & 1105897 & 27.109 & 5.003 & 18.895 & 1542 \\
XX & 7524031 & 2.23, 35.97 & 1079020 & 28.444 & 5.928 & 19.575 & 1662 \\
YY & 7355950 & 2.46, 33.57 & 1061361 & 29.236 & 5.968 & 18.394 & 2028 \\
XY & 2832144 & 2.22, 37.66 & 124943 & 97.250 & 17.986 & 29.752 & 7216 \\
YX & 2316772 & 1.91, 36.13 & 443614 & 45.499 & 8.866 & 5.533 & 3386 \\
\enddata
\tablecomments{
The third column gives the percentages of hits falling in the frequency ranges of civil aviation and global navigation satellite system (GNSS), respectively. 
The columns No off hit, Unclustered, and Drift range give the percentages of events satisfying the no off-hit counterpart condition, not being assigned to any large off-hits populated cluster, and the date-dependent adopted drift rate range, respectively. 
}
\end{deluxetable}

\begin{figure}[htpb]
  \centering
  \includegraphics[width=0.95\textwidth]{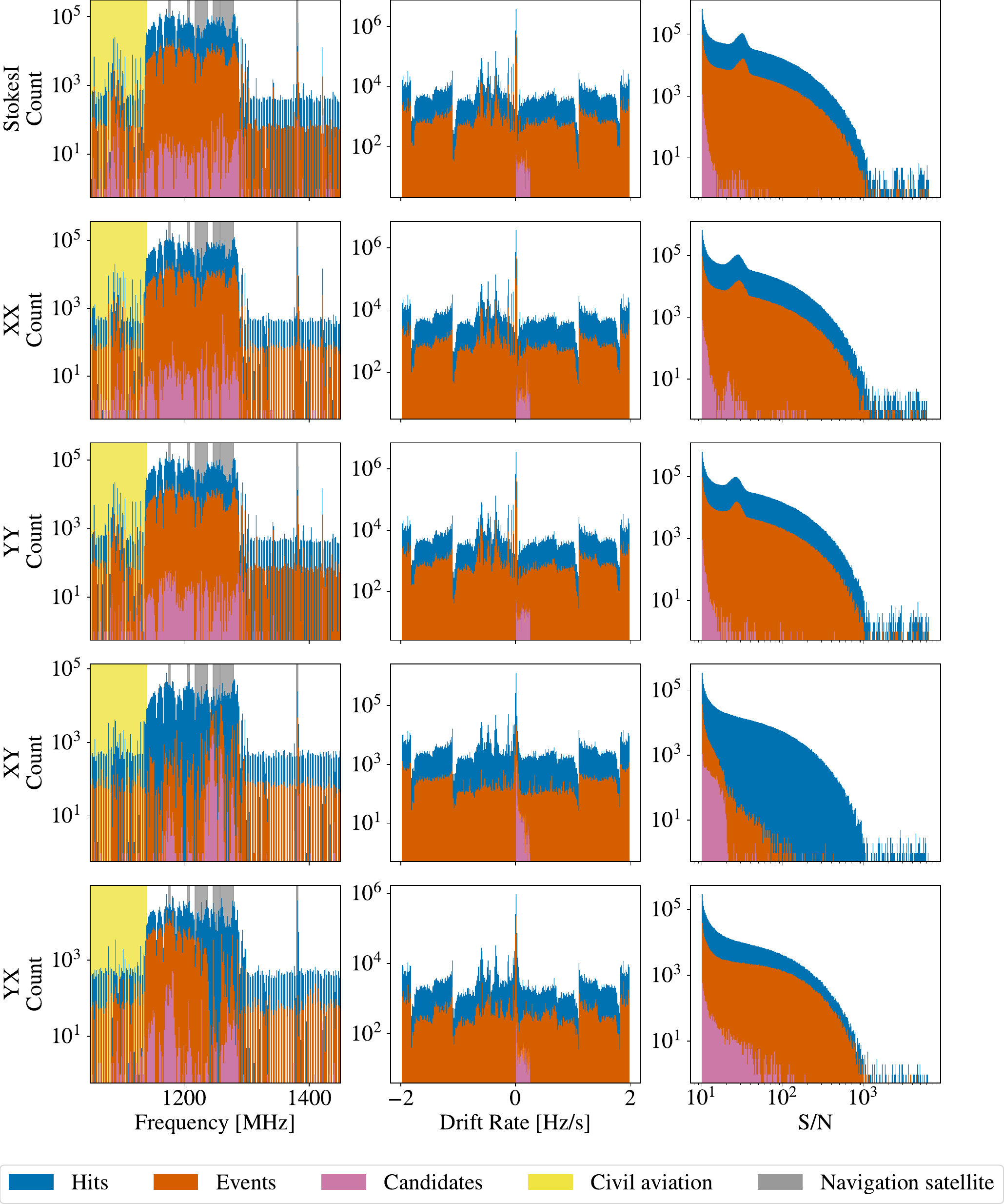}
  \caption{\label{fig:histograms_4pol}  Histograms of the distributions of frequency (left column), drift rate (middle column) and signal-to-noise ratio (right column) for the signals in total intensity and four polarization channels. }
\end{figure}

The distributions of hits and events in frequency, drift rate, and signal-to-noise ratio for total intensity and each polarization channel are illustrated in Figure \ref{fig:histograms_4pol}, and some known RFI sources within the observation frequency ranges \citep{2021RAA....21...18W} are also shown in the plots. The slight negative drift rate bias is caused by the downward relative acceleration of non-geosynchronous satellites \cite{2006JNav...59..293Z}. The signal population differences in polarizations are also reflected in the divergent distributions between XY and other polarization channels.

Using Stokes-$I$ as the reference catalog, we carry out a match for frequency counterparts in XX, YY, XY, and YX within one fine channel $\delta\nu_{\rm ch}$. The matched results are listed in Table \ref{table:pol_match}. The matched fraction is high at the hit level, with $94.7\%$ of Stokes-$I$ hits having a counterpart in at least one polarization product, and remains substantial at the event level ($82.3\%$). However, the matched fraction drops to $7.7\%$ for the potential candidate level. We also note that the matched fractions differ among the polarization products: the auto-correlation products XX and YY show higher correspondence with Stokes-$I$ than the cross-correlation products XY and YX. This suggests that a signal identified in Stokes-$I$ or in one polarization product may not always appear as a detected counterpart in every other polarization product, possibly because of differences in apparent signal power, noise properties, and the response of the search and filtering algorithm across polarization channels.

\begin{deluxetable}{lccccc}
\tablecaption{\label{table:pol_match}Stokes-$I$ referenced frequency matching with polarization products.}
\tablehead{
\colhead{Catalog} &
\colhead{XX (\%)} &
\colhead{YY (\%)} &
\colhead{XY (\%)} &
\colhead{YX (\%)} &
\colhead{Any pol. channel (\%)}
}
\startdata
Hits       & 86.7 & 86.7 & 63.4 & 60.7 & 94.7 \\
Events     & 73.5 & 71.9 & 3.5  & 28.9 & 82.3 \\
Potential Candidates & 4.9  & 2.5  & 0.1  & 0.8  & 7.7  \\
\enddata
\tablecomments{
Stokes-$I$ is used as the reference catalog. A match is defined by a frequency separation smaller than one channel width $\delta\nu_{\rm ch}=7.45\,{\rm Hz}$.
}
\end{deluxetable}

\begin{figure}[htpb]
  \centering
  \includegraphics[width=0.85\textwidth]{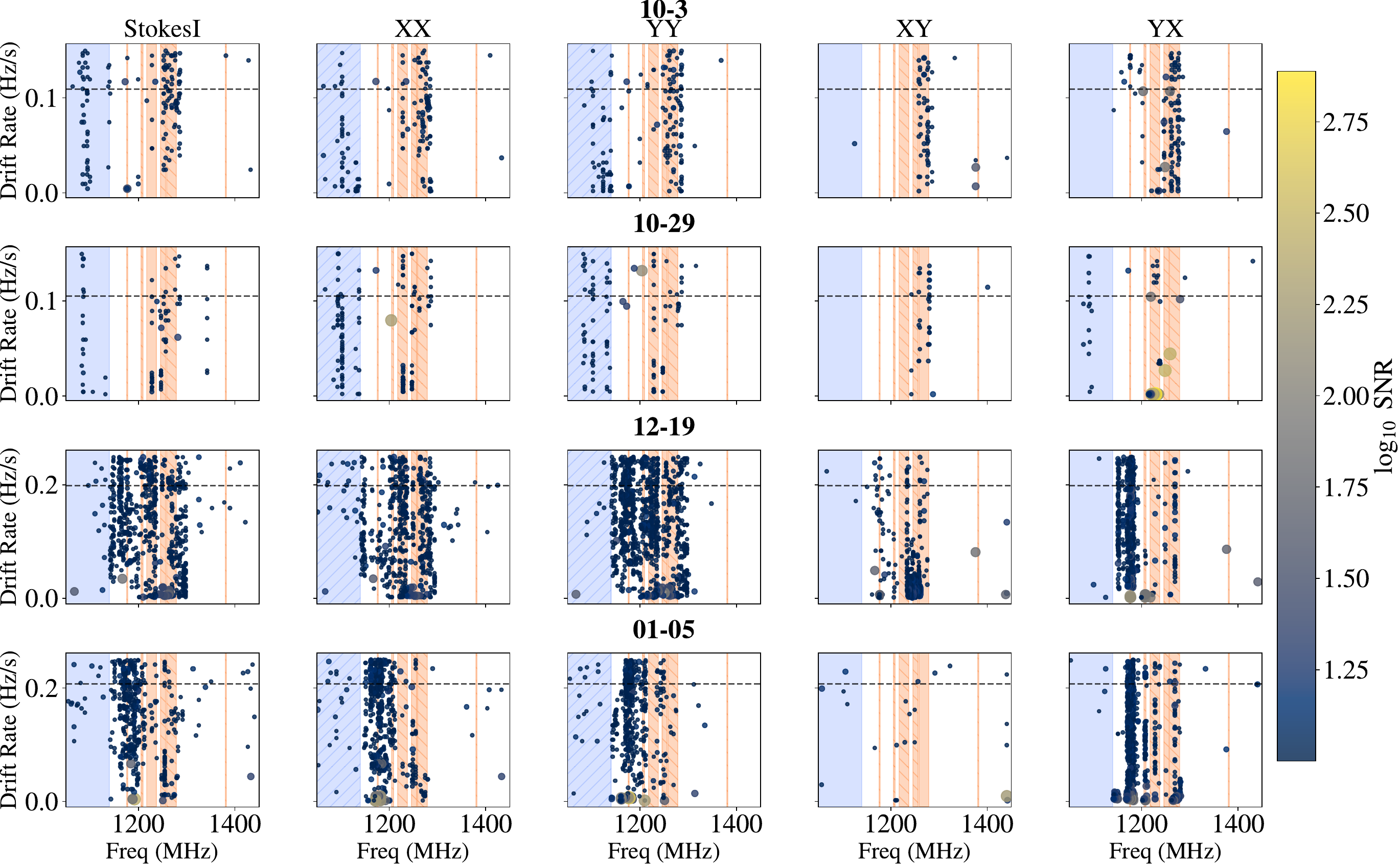}
  \caption{\label{fig:candidate_scatter}  Scatter plots of potential candidate in frequency, drift rate and signal-to-noise ratio for total intensity and four polarization channels. The color and the size for each point represent the value of SNR in log space. The blue and ornage color bar are the frequency ranges of civil aviation and navigation satellites, respectively. The black dashed lines represent the expected drift rates for each observation date.}
\end{figure}

We further investigate the distribution of potential candidates by the scatter plot in Figure \ref{fig:candidate_scatter}. A large fraction of potential candidates cluster within several restricted frequency windows rather than being uniformly distributed across the band. In particular, prominent concentrations coincide with the civil aviation band and the navigation satellites. The drift rates span a broad range, forming an overall band-like distribution, suggesting that these potential candidates are likely dominated by RFI with similar phenomenology.

\begin{figure}[htpb]
  \centering
  \gridline{
  \fig{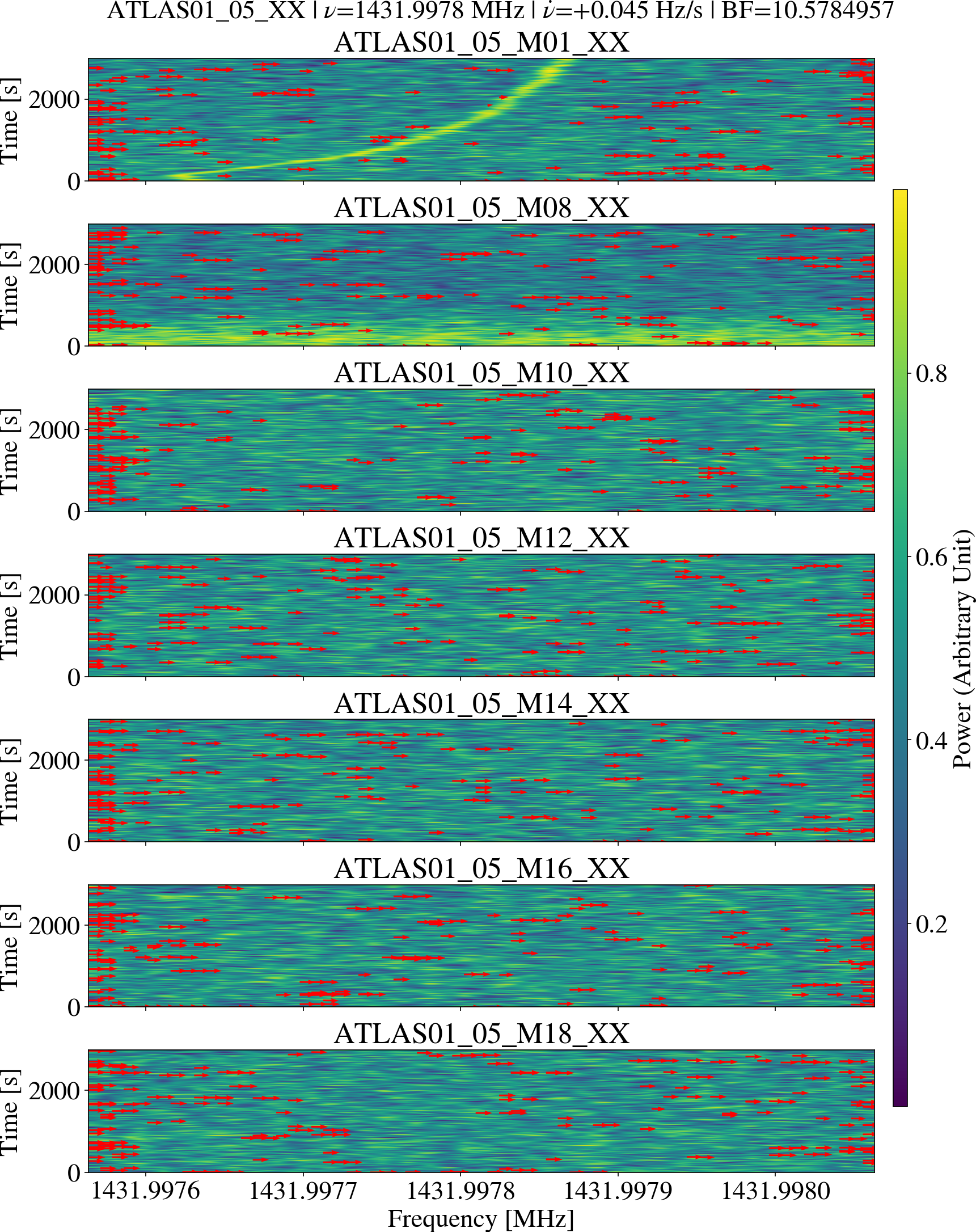}{0.45\textwidth}{(a)}
  \fig{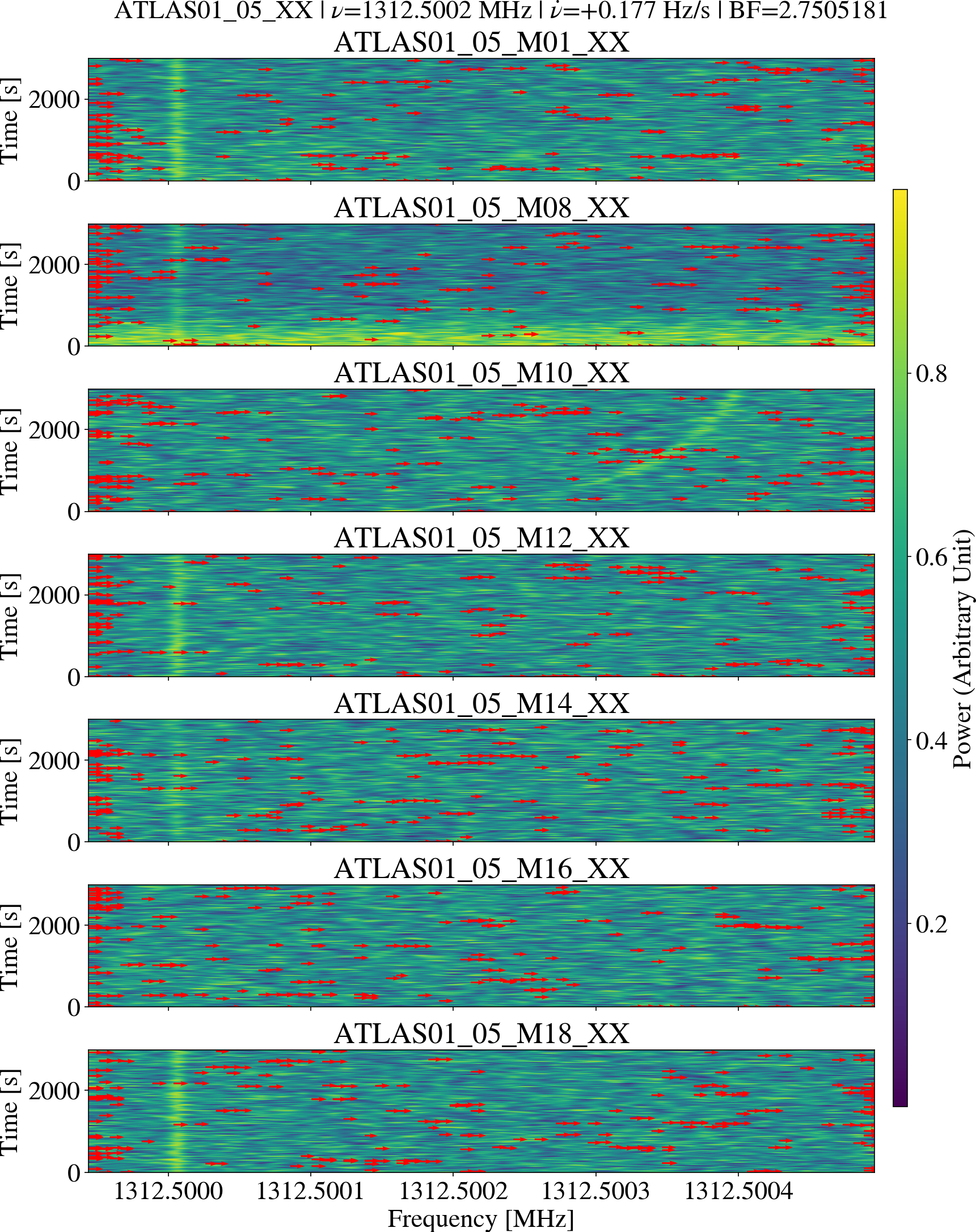}{0.45\textwidth}{(b)}
}
  \caption{\label{fig:chirp-up_signal}  (a) A chirp-up signal at frequency 1431.9978 MHz, with a drift rate of 0.045 $\mathrm{Hz \, s^{-1}}$ given by \texttt{bliss}. The chirp-up signal only appears in the central beam, while no obvious corresponding drifting signals appear in any reference beam. (b) A chirp-up signal at 1312.5502 MHz absent in the central beam, but appears in one of the reference beams. }
\end{figure}

\begin{figure}[htpb]
  \centering
  \includegraphics[width=0.85\textwidth]{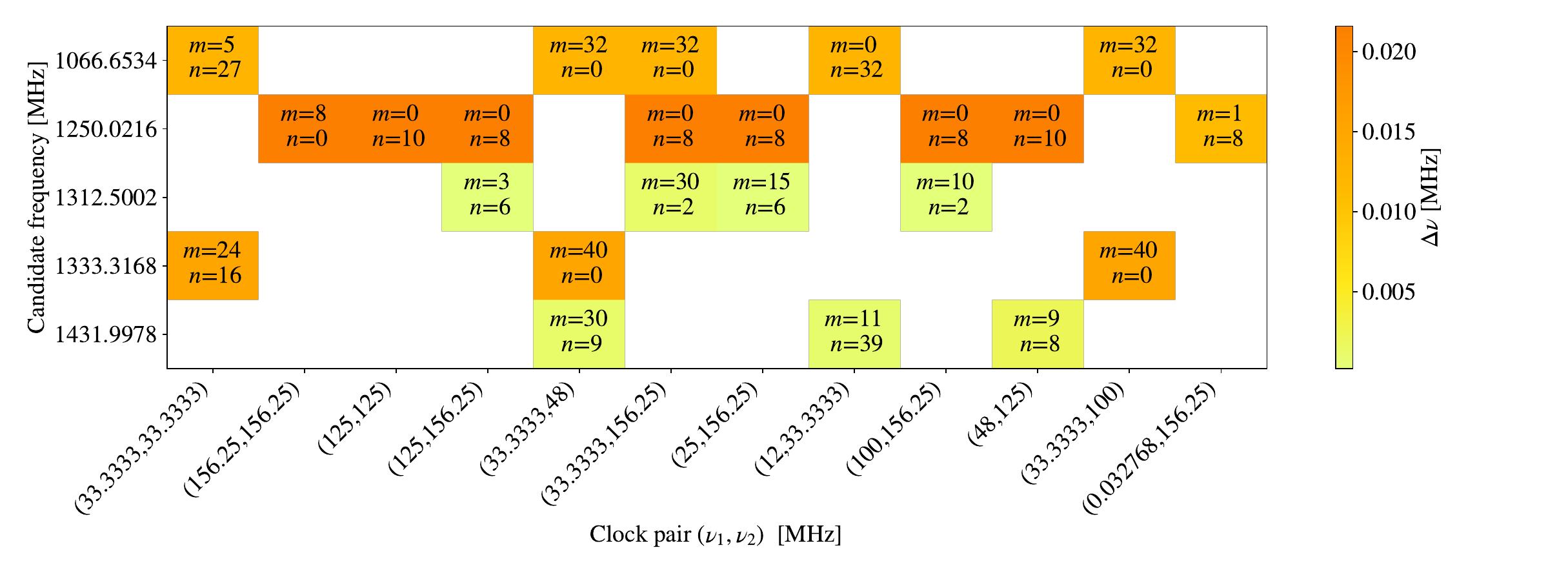}
  \caption{\label{fig:clock_oscillators_freq_pair}  Intermodulation analysis for the chirp-up signal frequencies. The horizontal axis lists the clock-frequency pairs $(\nu_i,\nu_j)$, and each row corresponds to one signal frequency $\nu_{\rm obs}$. Each cell reports the integers $(m,n)$ of the best match $\nu_{\rm model}=m\nu_1+n\nu_2$ (with $m,n\ge 0$ and $m+n\leqslant 50$). The colormap shows the absolute residual $|\nu_{\rm model}-\nu_{\rm obs}|$.}
\end{figure}

All selected potential candidates are reexamined by a visual inspection of the dynamic spectra. Almost all the potential candidates are two kinds of obvious false positives mentioned in \cite{2022AJ....164..160T}, which can be directly excluded. Specifically, those potential candidates fall at the expected drift rates in Figure \ref{fig:candidate_scatter} are also filtered during the visual inspection. Seven potential candidates survive after visual inspection, and exhibit similar ``chirp-up'' feature, instead of constant linear drift, which are listed in Table \ref{table:cand_list}. All these seven chirp-up signal are found in the observation on 5 January 2026.
\begin{deluxetable}{lcccc}[htpb]
\tablecaption{Frequencies and drift rates of the seven chirp-up potential candidates.\label{table:cand_list}}
\tablehead{
\colhead{Event ID} & 
\colhead{Central Frequency [MHz]} &
\colhead{Drift rate [$\mathrm{Hz \, s^{-1}}$]} &
\colhead{Polarization channel} &
\colhead{ln BF}
}
\startdata
244 & 1431.9978 & 0.045 & StokesI  & 2.423\\
250 & 1431.9978  & 0.045 & XX & 2.359\\
3703 & 1250.0126 & 0.055 & YX  & -2.035\\
5965 & 1333.3168 & 0.135  & YY  & 0.768 \\
47992 & 1250.0126 & 0.047 & YY  & -0.199 \\
225671 & 1066.6534 & 0.115 & YY & 1.097 \\
250224 & 1066.6534 & 0.107 & StokesI & -1.087\\
\enddata
\end{deluxetable} 
Figure \ref{fig:chirp-up_signal} (a) illustrates an example of these chirp-up drifting signals (Other six chirp-up potential candidates are shown in Appendix \ref{sec:appendix_Chirp_up}). We also identify a potential candidate exhibiting a chirp-up morphology, however, a morphologically similar signal is present in the reference beam (See Figure \ref{fig:chirp-up_signal} (b)). We therefore reject this as RFI, which suggests that the chirp-up signals may share a common instrumental origin. 
Motivated by this, we investigate whether the observed chirp-up frequencies can be explained as intermodulation products of the nominal clock-oscillator frequencies used on the ROACH 2 FPGA board.\footnote{The nominal frequencies of all crystal oscillators used on ROACH2 are listed at \url{https://github.com/ska-sa/roach2_hardware/blob/master/release/rev2/A/BOM/ROACH-2_sEV2_BOM.csv}.} Specifically, we consider integer linear combinations of the form $\nu = m\nu_1 + n\nu_2$, where $\nu_1$ and $\nu_2$ are candidate oscillator frequencies and $m$ and $n$ are non-negative integers. By scanning over $(\nu_1,\nu_2,m,n)$, we construct intermodulation sequences and assess whether they can account for the measured frequencies of the chirp-up signal. Figure \ref{fig:clock_oscillators_freq_pair} summarizes the best-matching pairings for these signals. All five frequencies admits one or more such explanations within absolute residual $|\nu_{\rm model}-\nu_{\rm obs}|\leqslant 0.03 {\rm MHz}$. Moreover, the solution sets overlap across different frequencies in terms of the oscillator frequency pairs, indicating that the seven signals can be linked through a common clock oscillator frequency library. This behavior is consistent with an instrumental (clock-related intermodulation) RFI origin. Additionally, since there is currently no evidence indicating any abnormal acceleration variations in the 3I/ATLAS, and the drift rates of these signals do not align with the expected drift rate for 5 January 2026, finally, we are inclined to classify these signals as RFI.

\section{Discussion}\label{sec:Discussion}
\subsection{Sensitivity}\label{subsec:Sensitivity}
The sensitivity of a radio observation can be determined by system equivalent flux density \citep{2013tra..book.....W,2017isra.book.....T}
\begin{equation}
  \mathrm{SEFD}=\frac{2k_\mathrm{B}T_\mathrm{sys}}{A_\mathrm{eff}},
  \label{SEFD}
\end{equation}
where $k_\mathrm{B}$ is the Boltzmann constant, $T_\mathrm{sys}$ is the system temperature, and $A_\mathrm{eff}$ is the effective collecting area. The sensitivity $A_\mathrm{eff}/T_\mathrm{sys}$ of FAST L-band 19 beam receiver is $\sim 2000 \, \mathrm{m^{2} \, K^{-1}}$ \citep{2011IJMPD..20..989N,2016RaSc...51.1060L,2019SCPMA..6259502J}. For narrowband signal detection (i.e., the signal bandwidth is narrower or equal to the observing spectral resolution), the minimum detectable flux density $S_{\min}$ can be given by \citep{2017ApJ...849..104E,2021AJ....162...33G,2025arXiv251218142S}
\begin{equation}
  S_{\min}=\mathrm{SNR}_{\rm th}\frac{\mathrm{SEFD}}{\sqrt{\beta} \nu_t}\sqrt{\frac{\delta \nu_{\mathrm{ch}}}{n_\mathrm{pol}\tau_\mathrm{obs}}},
  \label{S_min}
\end{equation}
where $\tau_\mathrm{obs}$ is the effective observing duration, $\delta \nu_{\mathrm{ch}}$ is the frequency channel bandwidth, $\delta\nu_t$ is the transmitted signal width (we assume to be $\delta\nu_t \sim 1$ Hz), $n_\mathrm{pol}$ represents the number of polarization channels of the telescope and $\beta$ is the dechirping efficiency expressed as \citep{2021AJ....162...33G}\footnote{\footnote{In the $S_{\min}$ expression of \citet{2021AJ....162...33G}, the dechirping-efficiency factor appears as $\beta$ in the denominator. For consistency with \texttt{bliss}, which recovers $\sqrt{\beta}$ of the signal power, we use the corresponding $\sqrt{\beta}$ factor in our calculation.
}}
\begin{equation}
  \beta=
  \begin{cases}
1 & \left\lvert \dot{\nu} \right\rvert \leqslant  \frac{\delta \nu_{\rm ch}}{\delta t}\\
\dfrac{\delta \nu}{\left\lvert \dot{\nu} \right\rvert \delta t}  & \left\lvert \dot{\nu} \right\rvert > \frac{\delta \nu_{\rm ch}}{\delta t} .
\end{cases}
\end{equation} 
The minimum detectable flux density can be used to estimate the minimum luminosity detection threshold based on the distance of the source $d$, which can be quantified by equivalent isotropic radiated power (EIRP) of the antenna:
\begin{equation}
  \mathrm{EIRP}=4\pi d^2 S_{\min} \delta\nu_t.
  \label{EIRP}
\end{equation}
Based on the distances and our observation configurations, the EIRP is estimated to be $\sim 10^{-3}$ W for our observations. We also define an Earth-directed EIRP per steradian $L_{\Omega} \equiv {\rm EIRP}/\Omega_\oplus$ as a comparative indicator for a hypothetical Earth-pointing transmitter, where 
\begin{equation}
  \Omega_\oplus(d)=2\pi\left(1-\cos\alpha\right), \quad \sin\alpha = \frac{R_\oplus}{d}
\end{equation}
is the solid angle corresponding to the Earth as seen from 3I/ATLAS, with $R_{\oplus}$ being the radius of Earth. The EIRP values and the Earth-directed EIRP values per steradian are listed in Table \ref{table:EIRP_list}. 
\begin{deluxetable}{lccc}[htpb]
\tablecaption{EIRP values, solid angle of Earth, and the Earth-directed EIRP per steradian of 3I/ATLAS during our observation dates.\label{table:EIRP_list}}
\tablehead{
\colhead{Observation Date} & 
\colhead{EIRP (W)} &
\colhead{$\Omega_{\oplus}$ (sr)} &
\colhead{$L_{\Omega}$ (W/sr)}
}
\startdata
2025-10-03 & $5.497\times10^{-3}$ & $9.140\times10^{-10}$ & $6.014\times10^{6}$\\
2025-10-29 & $4.727\times10^{-3}$ & $1.063\times10^{-9}$ & $4.446\times10^{6}$ \\
2025-12-19 & $2.862\times10^{-3}$ & $1.755\times10^{-9}$ & $1.631\times10^{6}$ \\
2026-01-05 & $3.349\times10^{-3}$ & $1.500\times10^{-9}$ & $2.233\times10^{6}$
\enddata
\end{deluxetable}

\subsection{Bayesian Limits on the Detection of Technosignatures}
We consider the hypothesis that a comet may host a narrowband radio transmitter with a small prior probability $p_T$. Let $T\in\{0,1\}$ denote latent indicator of whether the target hosts a narrowband transmitter $(T=1)$ or not $(T=0)$. For a radio transmitter with EIRP strength parameter $L>0$, the single-trial detection probability to detect the signal from the transmitter can be given by
\begin{equation}
  \mathcal{P}_{{\rm det}}(L)=\Phi(\sqrt{\frac{n_{\rm pol} \tau_{\rm obs}}{\delta\nu_{\rm ch}}}\frac{\sqrt{\beta} L}
{4\pi d^2{\rm SEFD}}  - {\rm SNR}_{\rm th})=\Phi(\zeta  L - {\rm SNR}_{\rm th}),
  \label{eq:P_det}
\end{equation}
where $\zeta$ collects the instrumental sensitivity, integration time, spectral resolution, and target distance dependence, $\Phi(x)$ is the standard normal cumulative distribution function. 

To account for the different frequency coverages of different surveys, we introduce a frequency coverage fraction. Let
$[\nu_{\rm ref,min},\nu_{\rm ref,max}]$ be the assumed reference emission frequency range of the transmitter, and let
$[\nu_{\min},\nu_{\max}]$ be the searched frequency range of the observation. The covered bandwidth within the reference range is
\begin{equation}
\Delta\nu_{\rm cov}
=
\max\left[
0,\,
\min(\nu_{\max},\nu_{\rm ref,max})
-
\max(\nu_{\min},\nu_{\rm ref,min})
\right].
\label{eq:dnu_cov}
\end{equation}
Assuming a uniform transmitter frequency distribution over $[\nu_{\rm ref,min},\nu_{\rm ref,max}]$, the frequency-coverage fraction is
\begin{equation}
f_{\nu}= \frac{\Delta\nu_{\rm cov}}{\Delta\nu_{\rm ref}}.
\label{eq:fnu_uniform}
\end{equation}
For the $k$-th observing configuration, we define the single-trial detection probability given $L$ as $\mathcal{P}_{{\rm det},k}(L)$. Under the assumption that trials are independent conditioned on $(T,L)$, the likelihood of observing all-zero outcomes given $T=1$ and $L$ is
\begin{equation}
\mathcal{L}(\mathrm{all0} \mid T=1,L)
=\prod_{k=1}^{K}[1-f_{\nu,k}\mathcal{P}_{{\rm det},k}(L)]^{N_k}.
\label{eq:like_all0_given_L}
\end{equation}
While for $T=0$, the likelihood is
\begin{equation}
\mathcal{L}(\mathrm{all0}\mid T=0)=1.
\label{eq:like_all0_T0}
\end{equation}
The joint posterior $p(p_T,L\mid \mathrm{all0})$ can be given by 
\begin{equation}
  p(p_T,L\mid \mathrm{all0})=
\frac{\mathcal{L}(\mathrm{all0}\mid T=1,L) \pi(L) p_T \pi(p_T)}
{p(\mathrm{all0})},
\label{eq:like_all0_joint}
\end{equation}
where $p(\mathrm{all0})$ is the evidence of all-zero observation result, $\pi(p_T)$ is the prior of $p_T$ and $\pi(L)$ is the prior of $L$. The details derivations for above equations are given in Appendix \ref{subsec:appendix_Posterior}. Apart from the all-zero result in this work, we also include the observation configurations of null results in ATA \citep{2025arXiv251218142S}, GBT \citep{2025RNAAS...9..351J} and MeerKAT \cite{2025ATel17499....1P} for joint constraints. The SEFD of ATA are fitted by the values for different center frequencies. And for simplicity, we assume that reference frequency range of the transmitter is 1-12 GHz, so that $\Delta \nu_{\rm cov}$ is equal to the frequency range for each observation.\footnote{The frequency range of the transmitter, as well as its distribution, can be also modeled by Bayesian metohds, but here, we just use this simple frequency coverage fraction for illustrative propose.}.

\subsubsection{Priors for Transmitter Occurrence Probability and EIRP}\label{subsubsec:Prior}

The analysis above requires specifying priors for the transmitter occurrence probability $p_T$ and the transmitter EIRP $L$. We adopt simple prior families designed to encode only two minimal expectations. The probability that a given target hosts an active narrowband transmitter is small $(p_T\ll 1)$, while its order of magnitude remains uncertain. Accordingly, we use flexible distributions on $p_T$ that concentrate most prior mass at low values yet do not enforce a single characteristic scale. Specifically, we explore two alternative choices for $p_T$: (i) a scaled-beta prior, in which $q\sim \mathrm{Beta}(a,b)$ and $p_T = p_{T,\max} q$, thereby enforcing $0<p_T\le p_{T,\max}$ while allowing the concentration toward small $p_T$ to be tuned by $(a,b)$; and (ii) a logit-normal prior, in which $z \sim \mathcal{N}(\mu_p,\sigma_p^2)$ and $ p_T = p_{T,\max}\,\mathrm{sigmoid}(z)$, equivalently representing a logit-normal prior rescaled to $(0,p_{T,\max})$. The EIRP of the transmitter on the comet is not expected to be extreme, but it may plausibly span a relatively small orders of magnitude, this motivates placing a log-normal prior $\ln L \sim \mathcal{N}(\mu_L,\sigma_L^2)$, so that multiplicative uncertainty is represented naturally. These assumptions are deliberately conservative and are used to illustrate how non-detections map to constraints on $(p_T,L)$. 

Because the observational outcome is an all-zero non-detection, we report one-sided Bayesian credible upper limits on the transmitter occurrence probability from the marginal posterior of $p_T$. The EIRP parameter $L$ is retained in the Bayesian model as an assumed transmitter power scale, but it is treated as a nuisance parameter and marginalized over from the joint posterior in Eq.~(\ref{eq:like_all0_joint}). The 95\% upper limit $p_{T,95}$ is defined by
\begin{equation}
\int_0^{p_{T,95}} p(p_T\mid{\rm all0})\,dp_T = 0.95 .
\end{equation}

With the adopted fixed parameter choices listed in Appendix~\ref{subsec:appendix_Parameters}, the two single-layer prior choices give similar illustrative upper limits on $p_T$. For the logit-normal prior, we obtain $\log_{10}p_{T,95}=-4.720$, while for the scaled-beta prior we obtain $\log_{10}p_{T,95}=-4.419$. These values are nearly unchanged among the individual observation configurations in this fixed-parameter single-layer calculation. The primary information of all-zero result is from Eq.~(\ref{eq:like_all0_given_L}). When $p_T\ll1$ and/or $f_{\nu}\mathcal{P}_{{\rm det}}(L)$ remains small over much of the prior-supported range, the likelihood provides only limited leverage on $p_T$ after marginalizing over $L$. Therefore, in this fixed-parameter single-layer demonstration, the marginal upper limits are only weakly differentiated by the individual observing configurations and should be interpreted together with the adopted prior choices.

\subsubsection{Hierarchical Prior Models}\label{subsubsec:HierPrior}
To avoid over-committing to a single, fixed prior scale for the transmitter occurrence probability in the weak-information (all-zero) regime, we adopt hierarchical prior families in which the distributional shape parameters are themselves treated as unknowns and inferred jointly with the occurrence probability. The EIRP parameter $L$ is retained as an assumed transmitter-strength scale and marginalized over when deriving the upper limit on $p_T$.

We implement two hierarchical variants consistent with the previous non-hierarchical priors: (i) hierarchical scaled-beta prior, in which 
\begin{equation}
q \mid m,\kappa \sim \mathrm{Beta}(m\kappa,(1-m)\kappa).
\end{equation}
Here $m\in(0,1)$ controls the prior mean of $q$ through the expected value $\mathbb{E}(q\mid m,\kappa)=m$, while $\kappa>0$ controls the prior concentration (or dispersion) around $m$, and larger $\kappa$ yields a more concentrated prior and smaller $\kappa$ yields a more diffuse prior. We complete the hierarchical model by assigning hyperpriors
\begin{equation}
m \sim \mathrm{Beta}(\alpha_m,\beta_m),\quad
\kappa \sim \mathrm{Gamma}(a_\kappa,b_\kappa),
\end{equation}
where the Gamma distribution is parameterized by shape $a_\kappa$ and rate $b_\kappa$. (ii) hierarchical logit-normal prior, in which 
\begin{equation}
z \mid \mu_p,\sigma_p \sim \mathcal{N}(\mu_p,\sigma_p^2),
\end{equation}
where $p_T=p_{T,\max}\,\mathrm{sigmoid}(z)$, and assign hyperpriors to $(\mu_p,\sigma_p)$,
\begin{equation}
\mu_p \sim \mathcal{N}(\mu_{p,0},\tau_{\mu_p}^2),
\quad
\sigma_p \sim \mathrm{HalfNormal}(s_{\sigma_p}).
\end{equation} 
In this hierarchical model, $\mu_p$ controls the typical scale of $p_T$ in logit space, while $\sigma_p$ controls the dispersion of $p_T$ implied by the prior. The half-normal hyperprior on $\sigma_p$ ensures positivity and allows the width of the logit-scale distribution to be inferred rather than fixed.

For the transmitter EIRP, we adopt a hierarchical log-normal family to represent multiplicative uncertainty over orders of magnitude, 
\begin{equation}
\ln L \mid \mu_L,\sigma_L \sim \mathcal{N}(\mu_L,\sigma_L^2),
\end{equation}
with hyperpriors on the location and scale parameters, 
\begin{equation}
  \mu_L \sim \mathcal{N}(\mu_{L,0},\tau_{\mu_L}^2),
\quad
\ln\sigma_L \sim \mathcal{N}(\ln\sigma_{L,0},\tau_{\ln\sigma_L}^2),
\end{equation}
which represents multiplicative uncertainty in $L$ naturally on a logarithmic scale. In the occurrence probability constraints reported below, $L$ and its hyperparameters are treated as nuisance quantities and marginalized over.

In the all-zero detection regime, the constraints of the likelihood are driven primarily by the transition of the effective detection probability $f_{\nu}\mathcal{P}_{\rm det}(L)$ from low to high values across the prior-supported range of $L$. The constraint is obtained from the one-dimensional marginal posterior $p(p_T\mid{\rm all0})$ after marginalizing over $L$ and all hyperparameters. The 95\% upper limits $p_{T,95}$ for the occurrence probability are listed in Table~\ref{table:logp_T-logL}. The corresponding two-dimensional $\log_{10}p_T$--$\log_{10}L$ posterior contours are shown in Appendix~\ref{fig:Posterior_logpT_logL} as diagnostic visualizations of the remaining degeneracy between the occurrence probability and the assumed transmitter power scale, and of the sensitivity to the adopted prior family.

\begin{deluxetable}{lcc}[htpb]
\tablecaption{
One-sided 95\% Bayesian credible upper limits on the transmitter occurrence probability for the hierarchical prior models with $p_{T,\max}=0.005$. The EIRP parameter $L$ is treated as a nuisance parameter and marginalized over. These limits are conditional on the adopted prior families and hyperpriors.
\label{table:logp_T-logL}}
\tablehead{
\colhead{Observation} &
\colhead{$\log_{10} p_{T,95}$ (scaled-beta)} &
\colhead{$\log_{10} p_{T,95}$ (logit-normal)}
}
\startdata
ATA \citep{2025arXiv251218142S} & $-4.503$ & $-2.355$ \\
FAST (This work) & $-4.592$ & $-2.339$ \\
GBT \citep{2025RNAAS...9..351J} & $-4.376$ & $-2.341$ \\
MeerKAT \citep{2025ATel17499....1P} & $-4.428$ & $-2.355$ \\
All & $-4.443$ & $-2.356$ \\
\enddata
\end{deluxetable}

The upper limits in Table~\ref{table:logp_T-logL} are determined by both the observation configurations and the adopted hierarchical prior family. The former one is reflected in effective detection probabilities $f_{\nu,k}\mathcal{P}_{{\rm det},k}(L)$. The distance, sensitivity, threshold, and integration time determine the transition region of $\mathcal{P}_{{\rm det},k}(L)$ as a function of $L$, while the searched bandwidth enters through the frequency-coverage factor $f_{\nu,k}$. After marginalizing over the assumed EIRP scale $L$, these likelihood factors determine how strongly the all-zero result suppresses different values of $p_T$.

The scaled-beta and logit-normal hierarchical priors lead to noticeably different upper limits on $p_T$. This difference reflects the different ways in which the two prior families distribute probability mass over the allowed interval of $p_T$ once their shape parameters are allowed to vary. The scaled-beta model can place substantial probability density toward very small occurrence probabilities, leading to more stringent $p_{T,95}$ values, whereas the logit-normal model is more centrally distributed in logit space and therefore yields less stringent upper limits.

Finally, we emphasize that the purpose of these priors is not to claim a physically unique distribution for either $p_T$ or $L$, but to provide illustrative parameterizations demonstrating how non-detections map to make constraints on transmitter occurrence probability with the assumed transmitter EIRP.

\subsection{Events Featurization}\label{subsec:EventsFeaturization}
 
\begin{figure}[htpb]
  \centering
  \includegraphics[width=0.85\textwidth]{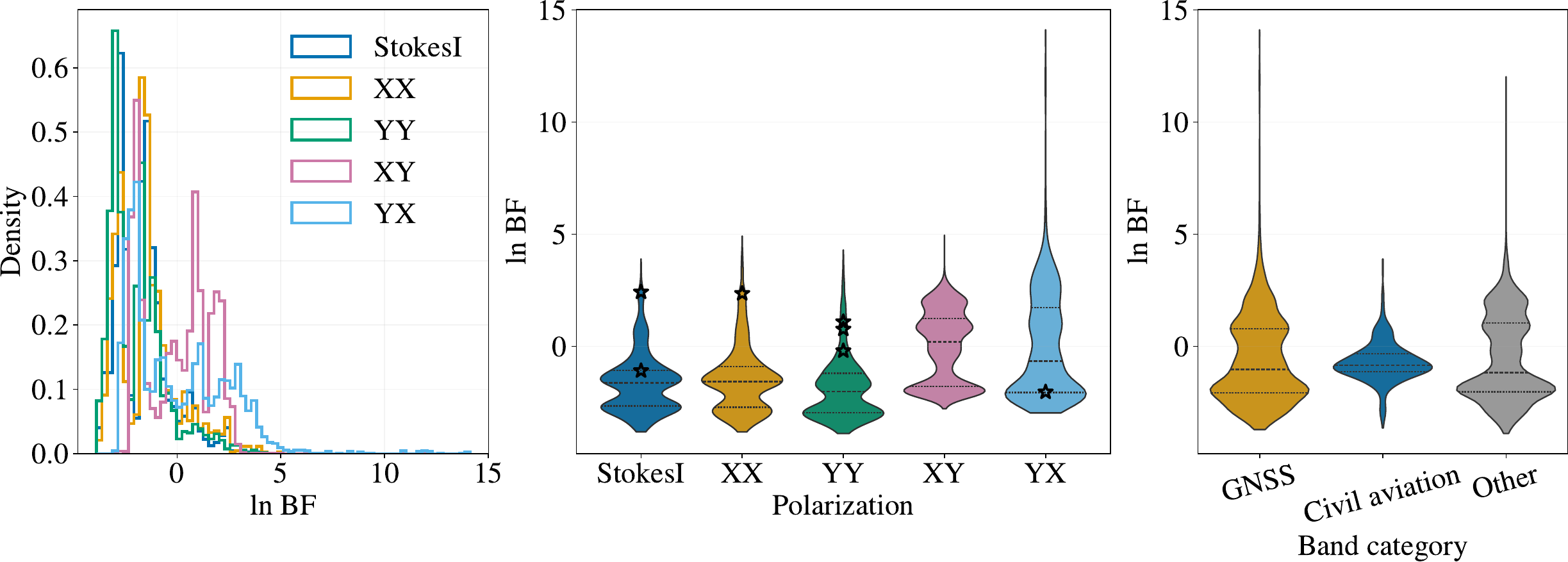}
  \caption{\label{fig:BF_summary} Distribution of the Bayes factor for potential candidates. Left: density histograms of for ln BF Stokes-$I$ and the four polarization products (XX, YY, XY, YX). Middle: violin plots of ln BF grouped by polarization product (dashed lines indicate quartiles), with black stars indicating the seven chirp-up potential candidates in Table \ref{table:cand_list}. Right: violin plots of ln BF grouped by frequency-band category.}
\end{figure}

Although all of the events are false positives, we still make statistical analysis for the Bayes Factor, coherence and PCA for them, so that we can investigate the characteristics for these false positives. We first summarize the empirical distributions of the Bayes factor in Figure \ref{fig:BF_summary} across polarization products and band categories. The total intensity and four polarization channels exhibit markedly different BF distributions: Stokes-$I$ as well as XX and YY are predominantly negative with modest tails, whereas the cross-polarization products (XY/YX) are broader and more multimodal, with a substantial fraction extending to ln BF>0. A similar dependence is observed across frequency categories: GNSS-rich regions show heavy-tailed, multimodal BF distributions, consistent with heterogeneous interference morphologies, while the civil-aviation band appears comparatively more homogeneous.

\begin{figure}[htpb]
  \centering
  \includegraphics[width=0.85\textwidth]{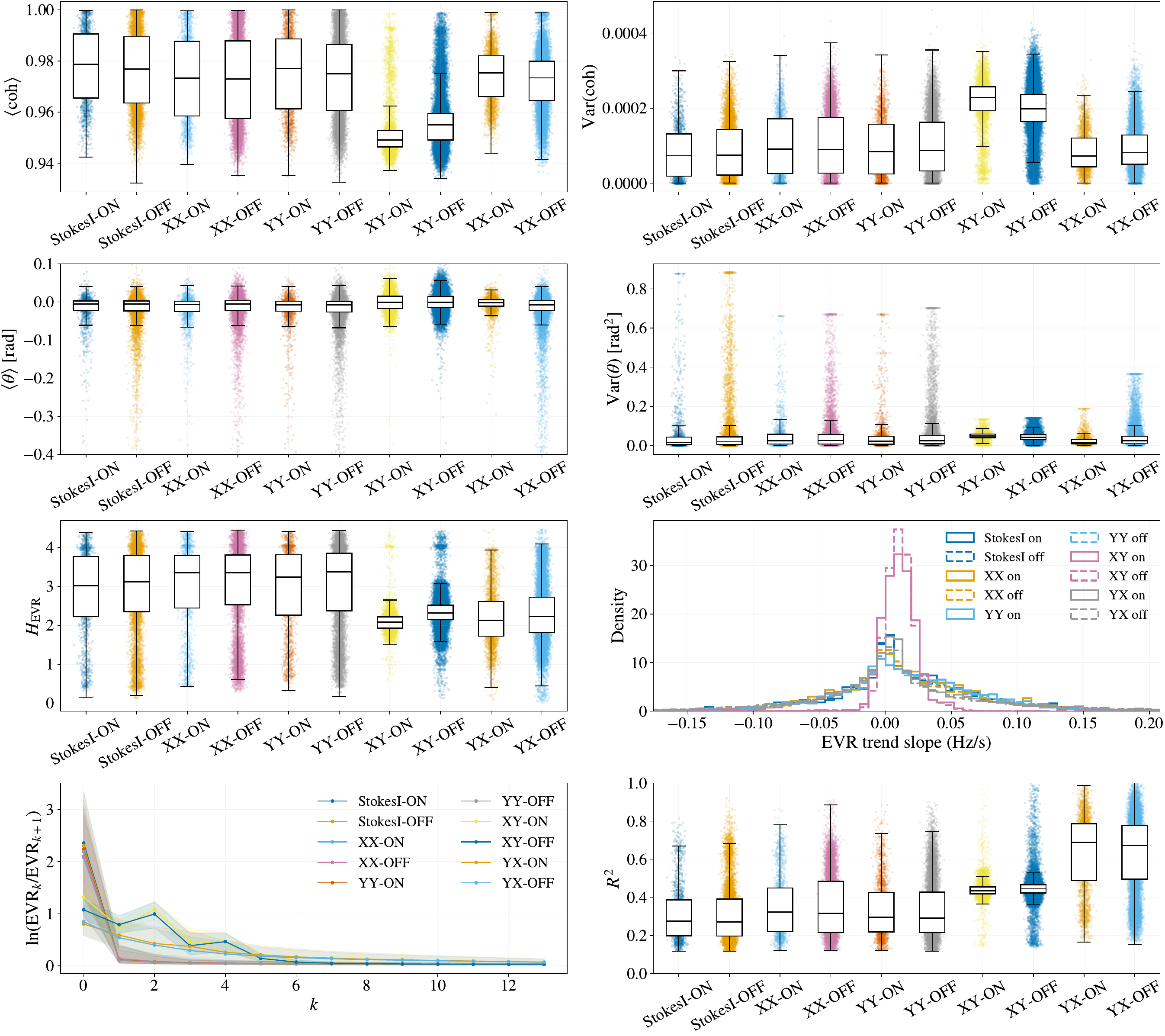}
  \caption{\label{fig:feature_stats_tensor_pca} Distributions of feature diagnostics derived from the structure tensor and PCA for the potential candidates in all beams, grouped by polarization channels and beams (ON: central beam; OFF: reference beams). First row: structure tensor statistics computed on the selected high-coherence regions, $\left\langle {\rm coh} \right\rangle $ and ${\rm Var \ (coh)}$. Second row: statistics of line orientation angle, $\left\langle \theta \right\rangle $ and ${\rm Var} \ \theta$ of the dynamic spectra. Third row: distributions of EVR entropy $H_{\rm EVR}$, and the EVR-weight trend slope $a_{EVR}$ of the dynamic spectra defined in Section \ref{subsubsec:PCA}. Fourth row: median trend of EVR ratio $\ln({\rm EVR}_k/{\rm EVR}_{k+1})$ versus $k$ for the top 15 components (curves show per-group medians and shaded regions show the interquartile range) and the coefficient of determination $R^2$ from fitting $\ln({\rm EVR}_m)$ in Eq. (\ref{eq:EVR}) across available components. Box plots summarize per-group distributions (median, IQR, and 1.5$\times$IQR whiskers) and colored points show individual samples in each group.}
\end{figure}

The statistics for the outputs resulting from structure tensor and PCA are illustrated in Figure \ref{fig:feature_stats_tensor_pca}. These features of the dynamic spectra in all beams of each event are stratified by polarization channel and by on/off beam. As we described in Section \ref{subsubsec:structure_tensor}, we select high-coherence regions to place the structure tensor quiver arrows. This restriction is intended to preferentially select locally anisotropic patches that, in an ideal narrowband drifting signal, would concentrate around the drift track and thereby highlight its linear geometry. If the selected region is dominated by a single, stable linear structure, $\mathrm{coh}$ within that region should be relatively uniform and $\mathrm{Var}(\mathrm{coh})$ should be small. Therefore, $\mathrm{Var}(\mathrm{coh})$ can be interpreted as a quality-control indicator for whether the selected high-coherence region behaves like a coherent structure instead aggregates multiple texture-driven patches. The line-orientation estimator $\theta_{\rm line}$ has group medians close to zero with modest dispersion, whereas ${\rm Var}(\theta)$ shows a long-tailed distribution, which is consistent with the characteristics of textured backgrounds or multi-component structures where no single drift-like ridge dominates the patch (See the examples of six event waterfall plots in Figure \ref{fig:chirp-up_signal} and Figure \ref{fig:four_chirp_up_signal}, almost all quiver arrows are horizontal without any significant inclination).

The EVR entropy $H_{\rm EVR}$ is systematically lower for the cross-hands (XY/YX) than for XX/YY, implying that the EVR distribution is more peaked toward the first few PCs, so that most of the variance can be represented by a smaller number of leading components with top rank. The EVR-weight trend slopes defined for rough frequency-drifting tendency, generally concentrated around a zero-drift mean values, especially the XY channel. The component-to-component ratios $\ln(\mathrm{EVR}_k/\mathrm{EVR}_{k+1})$ decrease rapidly with $k$ and approach a near-flat tail at larger $k$, showing that the PCA spectrum transitions from a steep leading part to a noise-dominated regime. Finally, the goodness-of-fit $R^2$ for the quadratic model $\ln(\mathrm{EVR}_m)$ varies substantially by group, with generally higher $R^2$ in the cross-hands products, indicating that those channels more often follow a smooth, rapidly decaying EVR envelope over the fitted range. 

Because the vast majority of the present sample consists of false positives and complex RFI or background noise, these statistics primarily reflect the background texture and instrumental/RFI imprint within the dynamic spectrum cutouts, rather than isolating the idealized signature of a single, clean Doppler-drifting narrowband track. A notable outcome is that the ON and OFF beams exhibit broadly similar population-level distributions for these features. This consistency is aligned with the fact that the vast majority of detections in the present sample are false positives: the dominant contribution to these statistics is therefore the morphology of background noise texture and persistent instrumental/RFI. At the same time, the nontrivial coherence levels and structured EVR patterns indicate that these RFI/noise realizations are not featureless. They often contain anisotropic and low-rank components that are quantified by the same diagnostics, and which must be accounted for when interpreting drifting-signal features.

\begin{figure}[htpb]
  \centering
  \gridline{
  \fig{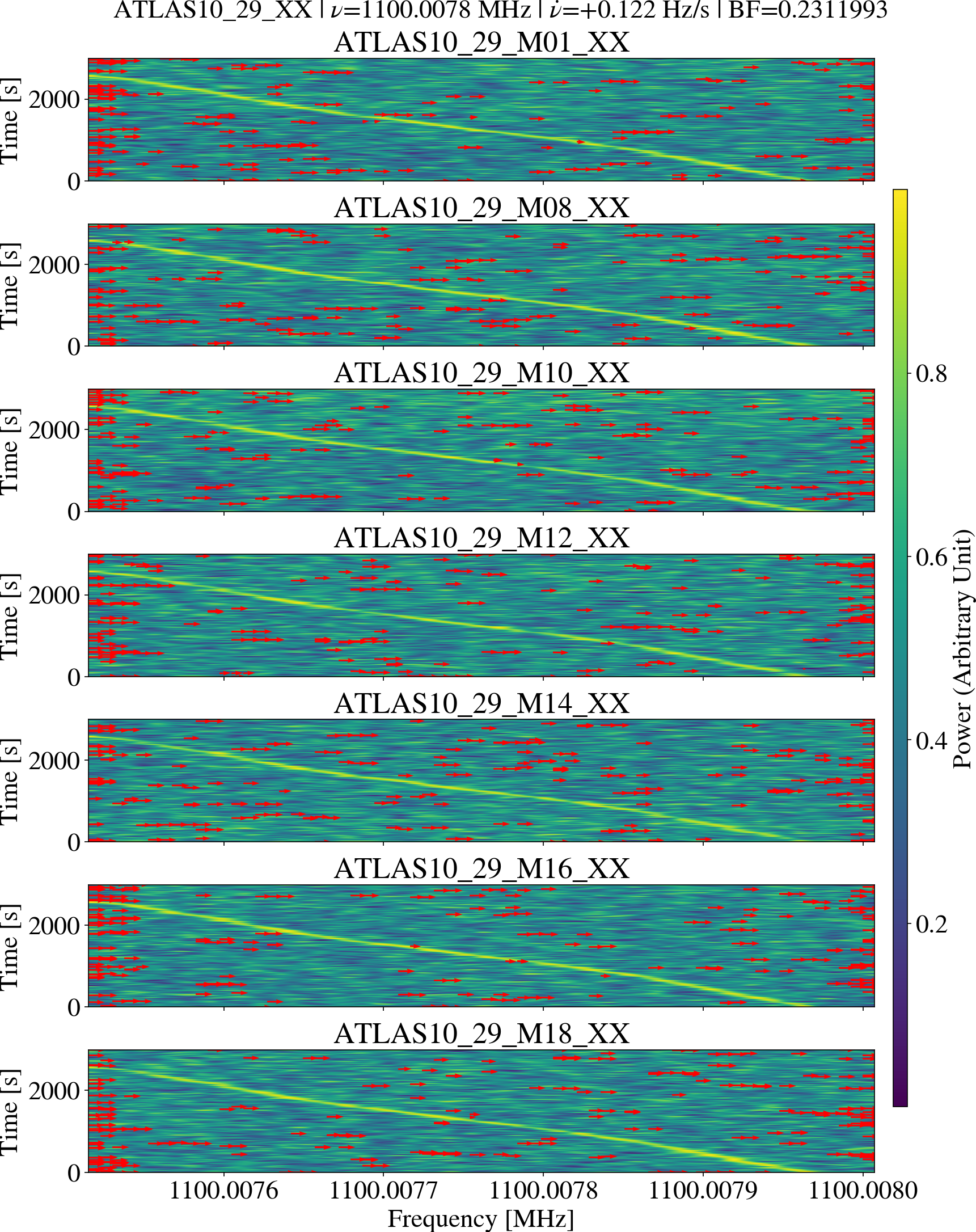}{0.34\textwidth}{}
  \fig{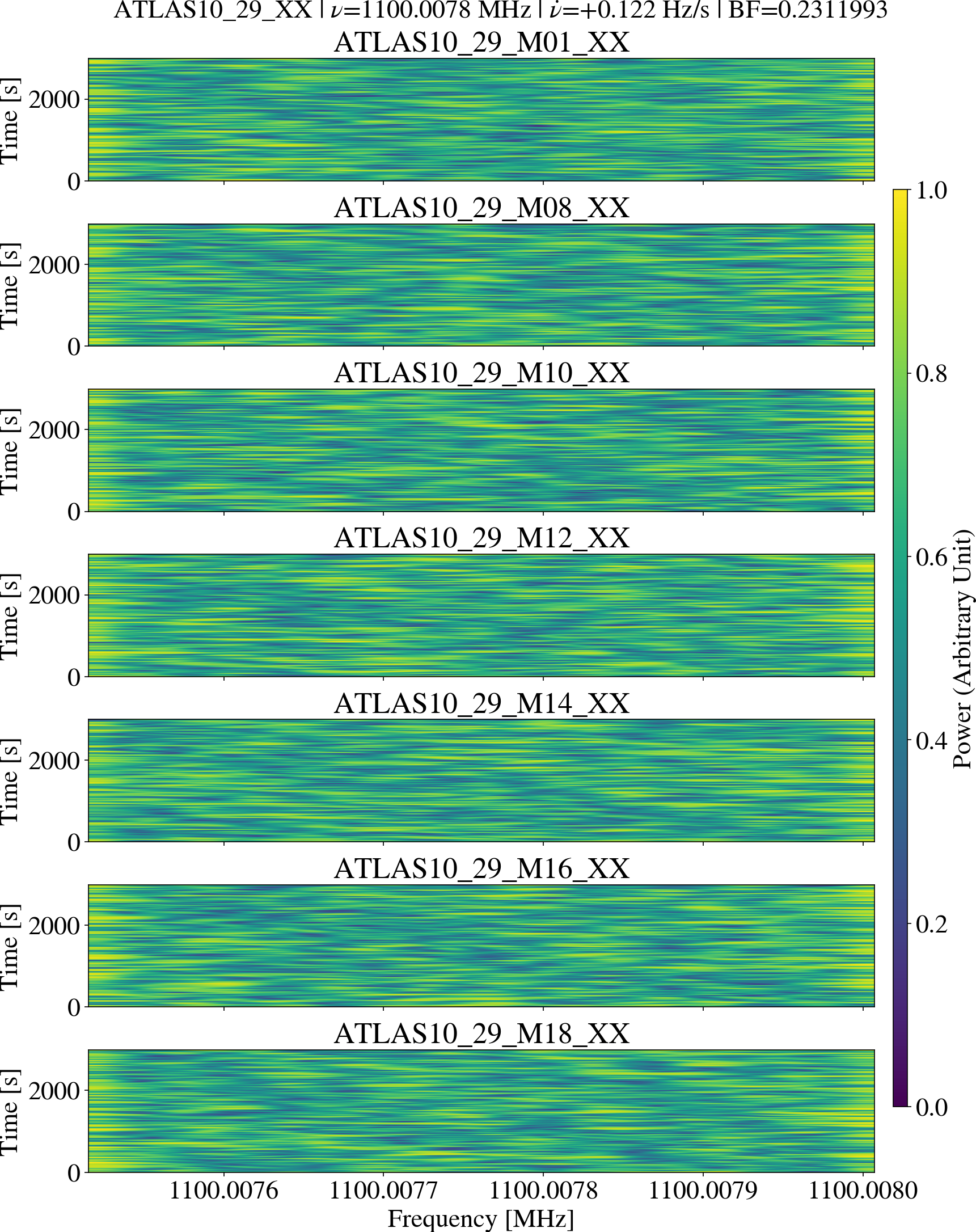}{0.34\textwidth}{}
  \fig{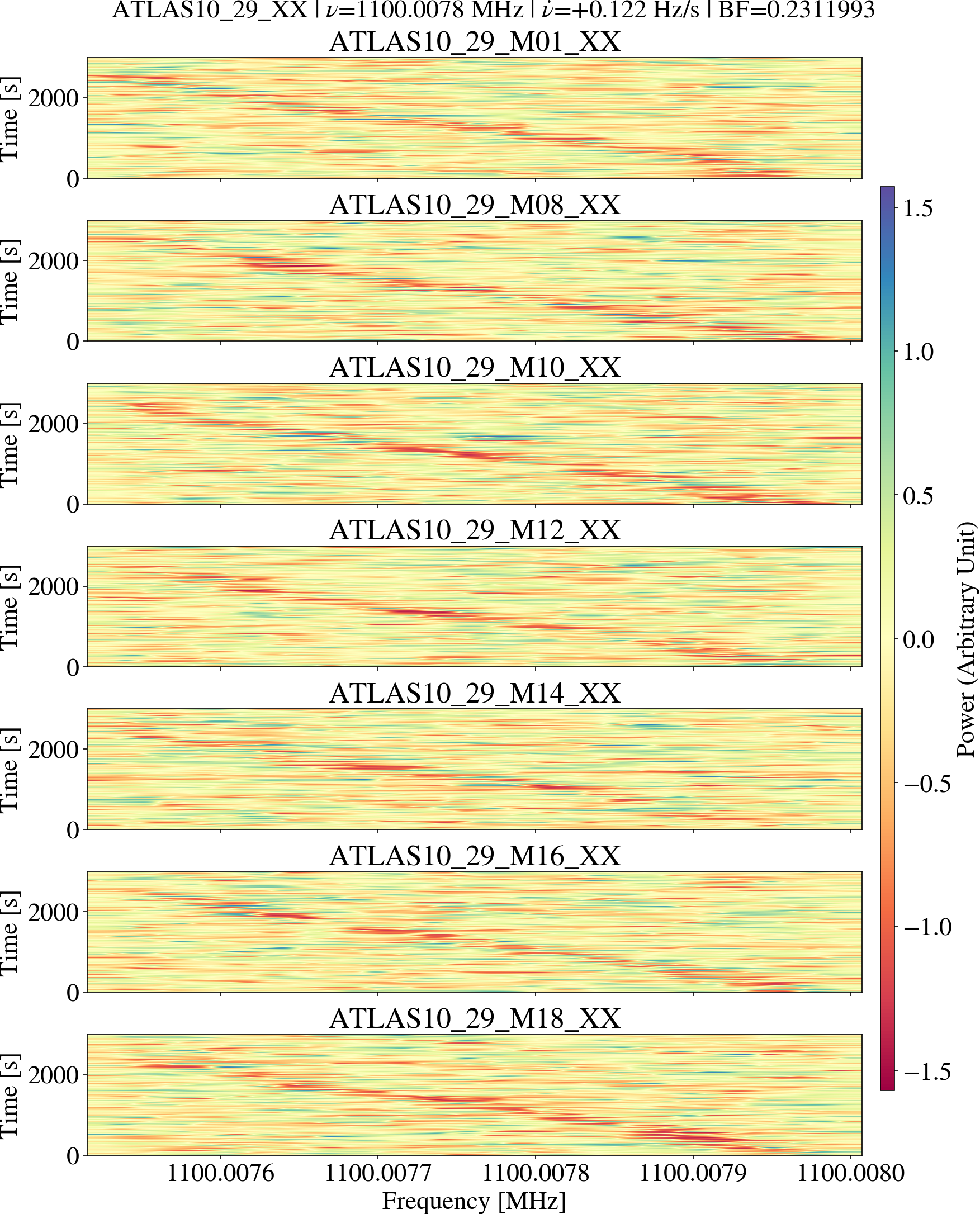}{0.34\textwidth}{}
}
\gridline{
  \fig{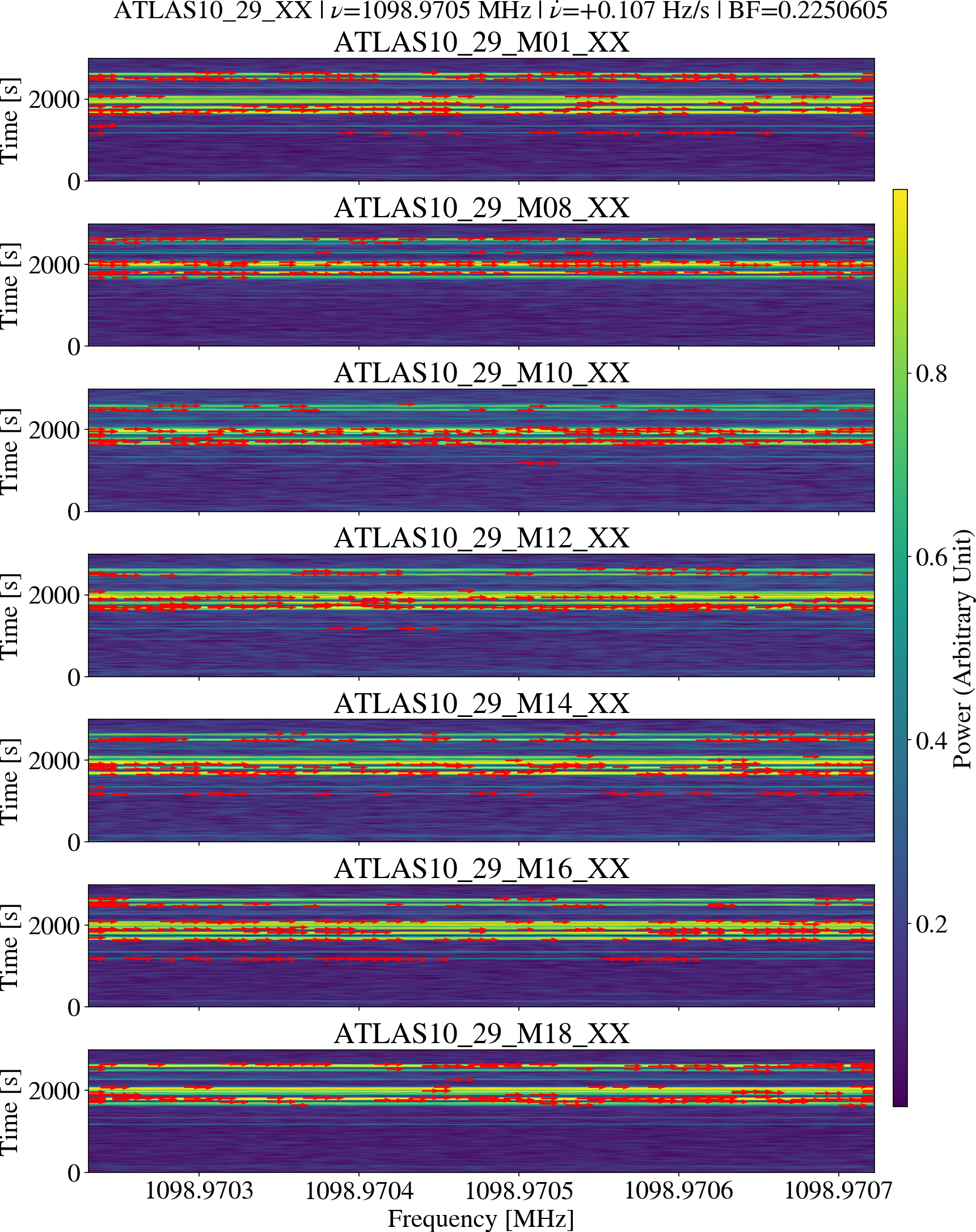}{0.34\textwidth}{}
  \fig{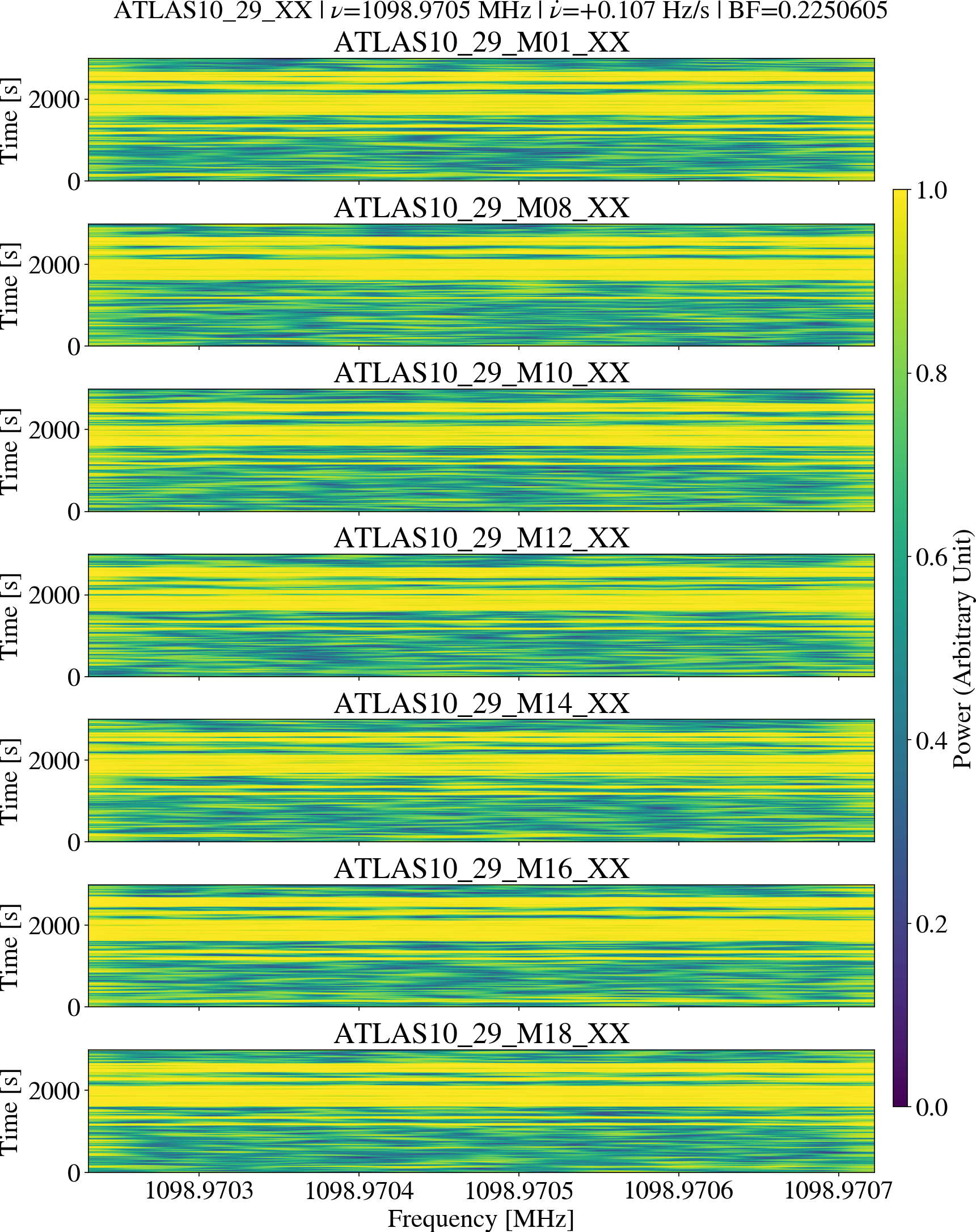}{0.34\textwidth}{}
  \fig{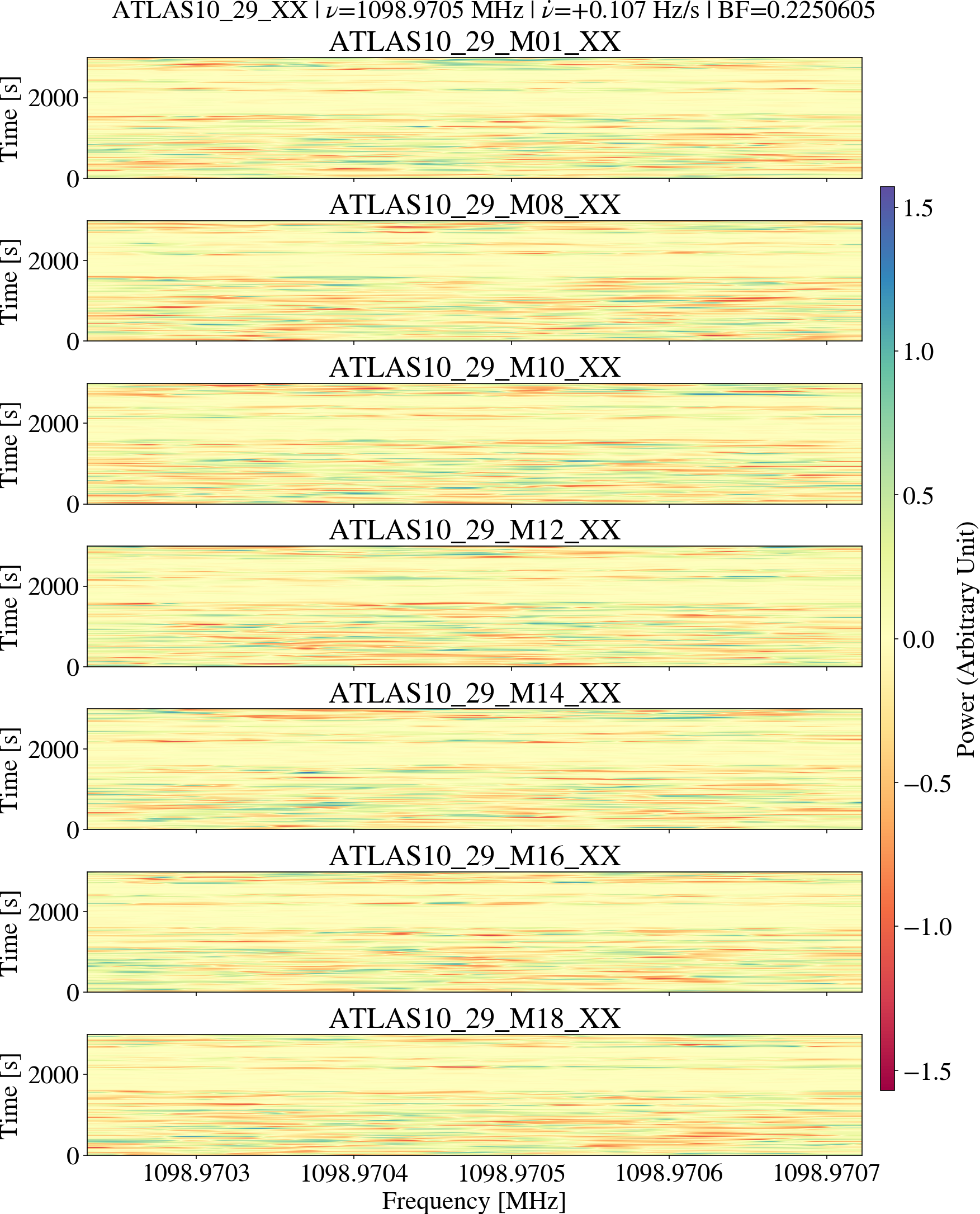}{0.34\textwidth}{}
}
  \caption{\label{fig:event_feature}  Waterfall plots (left), coherence map (middle) and orientation angle (right) for Doppler-drifting RFI signal (top) and relatively broadband RFI signal (bottom) appearing in all beams. }
\end{figure}

We also identify a plausible reason why visually apparent drifting tracks are not consistently traced by the structure-tensor quiver arrows. Coherence measure is derived from locally averaged second moments of the intensity gradients, and thus quantifies local anisotropy. In real dynamic spectrum, many visually apparent signal with drift-like features often show amplitude modulation and intermittency along the track, whereas the background often contains stable instrumental/RFI texture (e.g., quasi-horizontal striations). Under our high-coherence sampling strategy, the selected pixels preferentially trace the most repeatable anisotropic texture within each patch, which frequently aligns close to the frequency axis, yielding $\langle\theta\rangle$ concentrated near zero even when a drift-like pattern signal is present. Figure \ref{fig:event_feature} illustrates this behavior using two representative false-positive events that appear in all beams: one contains a visually clear linear drift, and the other is a relatively broadband horizontal feature. Although the drift is visible in both the waterfall and the orientation-angle map, its intermittency leads to spatially fragmented coherence, so the sampled quivers do not follow the track continuously. By contrast, the broadband horizontal feature is more uniform and produces a stronger, more spatially coherent anisotropic signature, and is therefore more reliably emphasized by the quiver arrows. 

\section{Conclusion}\label{sec:Conclusion}
We perform $\sim 4-{\rm hr}$ narrowband technosignature search toward 3I/ATLAS, the third confirmed ISO, with FAST L-band multibeam receiver in the frequency range of 1.05-1.45 GHz. The data was recorded on SETI backend with 7.5 Hz frequency resolution and 10 s sample time.  We search for narrowband drifting signal with drift rates consistent with the relative motion of 3I/ATLAS and SNR threshold above 10, in four polarization channels. 

The hit search is performed using \texttt{bliss} within $\pm \ 2 \ \mathrm{Hz \, s^{-1}}$. The obtained hits are filtered by HDBSCAN and expected drift rate. We apply Bayes factor to quantify the significance in the SNR differences between on-source and off-source. We also utilize structure tensor and PCA to featurize the morphology of the events.

After visual inspections, almost all potential candidates we obtain are obvious false positives and rejected as RFIs, except seven chirp-up signals detected on 5 January 2026, which only appear in the central beam. These seven chirp-up signals are finally rejected as instrumental RFIs instead of signal of interest, due to the consistency with the clock oscillator intermodulation in frequencies and inconsistency with the expected drift rate. Based on the null result, our search places an EIRP upper limit of $2.862\times10^{-3}\,\mathrm{W}$ for any narrowband radio transmitters. We also present a Bayesian model for deriving illustrative limits on the transmitter occurrence probability with assumed transmitter power distributions.

\begin{acknowledgments}
We sincerely appreciate the referee's suggestions, which help us greatly improve our manuscript. We thank the staffs in FAST data center for approving our DDT application on FAST observing time. We thank Chenoa Tremblay for sharing
the MeerKAT information about the technosginature search. This work was supported by National Key R\&D Program of China, No.2024YFA1611804 and the China Manned Space Program with grant No. CMS-CSST-2025-A01. This work made use of the data from FAST (Five-hundred-meter Aperture Spherical radio Telescope).  FAST is a Chinese national mega-science facility, operated by National Astronomical Observatories, Chinese Academy of Sciences.
\end{acknowledgments}


\appendix
\section{Explained Variance Ratio in PCA for Drifting Signal}\label{subsec:appendix_EVR_PCA}
Consider the dynamic spectrum $X_c\in\mathbb{R}^{N_t\times N_\nu}$. A narrow Doppler-drifting line can be idealized as a spectral profile,
\begin{equation}
X_c (t_i,\nu_j)\approx \alpha(t_i)g(\nu_j-\nu_0-\dot\nu,t_i),
\end{equation}
where $a(t_i)$ is the amplitude and $g(\nu)$ has characteristic width $w_\nu$. For drifting regime $|\dot{\nu}|\tau_{\rm obs}\gg w_\nu $, applying PCA to $X_c$ is equivalent to diagonalizing the frequency-domain covariance,
\begin{equation}
C_\nu \equiv \frac{1}{N_t}X_c^{T}X_c,\quad
\mathrm{EVR}_m=\frac{\lambda_m}{\sum_s \lambda_s},
\end{equation}
where $\{\lambda_m\}$ are the eigenvalues of $C_\nu$ (equivalently $\lambda_m=\sigma_m^2$, with $\sigma_m$ the singular values of $X_c$) Substituting the drifting-profile yields
\begin{equation}
C_\nu(\nu_j,\nu_k)=\frac{1}{N_t}\sum_{i=1}^{N_t}\alpha^2(t_i)
g\left(\nu_j-\nu_0-\dot{\nu}t_i\right)
g\left(\nu_k-\nu_0-\dot{\nu}t_i\right).
\label{eq:C_nu_app}
\end{equation}
When the drift spans many line widths, the correlation between channels depends primarily on the frequency separation $\Delta\nu=\nu_j-\nu_k$. Approximating $\alpha^2(t_i)$ by its time average $A^2\equiv \langle \alpha^2\rangle$ and replacing the discrete sum by an integral along the drift line gives an approximately shift-invariant kernel,
\begin{equation}
C_\nu(\nu_j,\nu_k)\approx K(\Delta\nu)
=\frac{A^2}{|\dot{\nu}|\tau_{\rm obs}}\int_{-\infty}^{\infty} g(u)g(u-\Delta\nu)du .
\end{equation}
For a Gaussian line profile $g(\nu)\propto \exp[-\nu^2/(2w_\nu^2)]$, the correlation kernel is also Gaussian,
\begin{equation}
K(\Delta\nu)= K_0 \exp\left(-\frac{\Delta\nu^2}{4w^2_\nu}\right).
\label{eq:kernel_gauss}
\end{equation}
Over the drift span, define the number of frequency channels traversed as
\begin{equation}
N_{\rm span}\equiv {\rm round}\left(\frac{|\dot{\nu}|\tau_{\rm obs}}{\delta\nu}\right),
\end{equation}
and consider the discrete Fourier modes
\begin{equation}
v_m(j)=\exp\left(i\frac{2\pi m j}{N_{\rm span}}\right),\qquad m=0,\dots,N_{\rm span}-1,
\end{equation}
which approximately diagonalize a shift-invariant covariance. The corresponding eigenvalues can be written as the discrete Fourier transform of the sampled kernel $K_n\equiv K(n\delta\nu)$,
For $|\dot\nu|\tau_{\rm obs}\gg \delta \nu$ case, 
\begin{equation}
\lambda_m=\sum_{n=0}^{N_{\rm span}-1}K_n\exp\left(-i\frac{2\pi m n}{N_{\rm span}}\right).
\end{equation}
In the large-span limit, this sum is well-approximated by the continuous Fourier transform
\begin{equation}
\lambda_m \approx \int_{-\infty}^{\infty} K(\Delta\nu)e^{-i \ell _m \Delta\nu}d(\Delta\nu),
\quad
\ell _m\equiv \frac{2\pi m}{|\dot{\nu}|\tau_{\rm obs}} .
\end{equation}
Using Equation (\ref{eq:kernel_gauss}) and the Gaussian Fourier transform,
\begin{equation}
\int_{-\infty}^\infty \exp\left(-\frac{x^2}{4w_\nu^2}\right)e^{-i\ell x}dx
=2\sqrt{\pi}w_\nu\exp\left[-(\ell w_\nu)^2\right],
\end{equation}
we obtain
\begin{equation}
\lambda_m \approx K_0(2\sqrt{\pi}w_\nu)\exp\left[-(\ell_m w_\nu)^2\right]
=K_0(2\sqrt{\pi}w_\nu)
\exp\left[-\left(\frac{2\pi m w_\nu}{|\dot{\nu}|\tau_{\rm obs}}\right)^2\right].
\label{eq:lambda_m_gauss}
\end{equation}

We define the EVR entropy as
\begin{equation}
H_{\rm EVR}=-\sum_m p_m\ln p_m,\qquad
p_m=\frac{\lambda_m}{\sum_s\lambda_s}.
\end{equation}
For $|\dot{\nu}|\tau_{\rm obs}\gg w_\nu$, define
\begin{equation}
a\equiv \left(\frac{2\pi w_\nu}{|\dot{\nu}|\tau_{\rm obs}}\right)^2\ll 1 .
\end{equation}
Up to normalization, Equation (\ref{eq:lambda_m_gauss}) implies $p_m\propto e^{-a m^2}$, so that
\begin{equation}
p_m \approx \frac{e^{-a m^2}}{\sum_s e^{-a s^2}}
\approx \sqrt{\frac{a}{\pi}}e^{-a m^2},
\end{equation}
where we have used the continuum approximation
\begin{equation}
\sum_s e^{-a s^2}\approx \int_{-\infty}^{\infty}e^{-a x^2}dx=\sqrt{\frac{\pi}{a}} .
\end{equation}
Treating $p(x)=\sqrt{a/\pi}\,e^{-a x^2}$ as a continuous density yields
\begin{equation}
H_{\rm EVR}\approx -\int_{-\infty}^{\infty} p(x)\ln p(x)dx
=\frac{1}{2}\ln\left(\frac{\pi e}{a}\right)
=\ln\left(\frac{|\dot{\nu}|\tau_{\rm obs}}{2\pi w_\nu}\right)+\frac{1}{2}\ln(\pi e).
\label{eq:Hevr_final}
\end{equation}

In addition, our derivation replaces the discrete modal sum and discrete Fourier transform by a continuous Gaussian integral and a continuous Fourier transform. For a finite drift span and a finite number of modes, this discrete-continuous approximation modifies the tail weights and induces spectral leakage in the discrete spectrum, which may increase $H_{\rm EVR}$ relative to the idealized expression.

\section{Chirp-up Drifting Potential Candidates}\label{sec:appendix_Chirp_up}
\begin{figure}[htpb]
\centering
\gridline{
  \fig{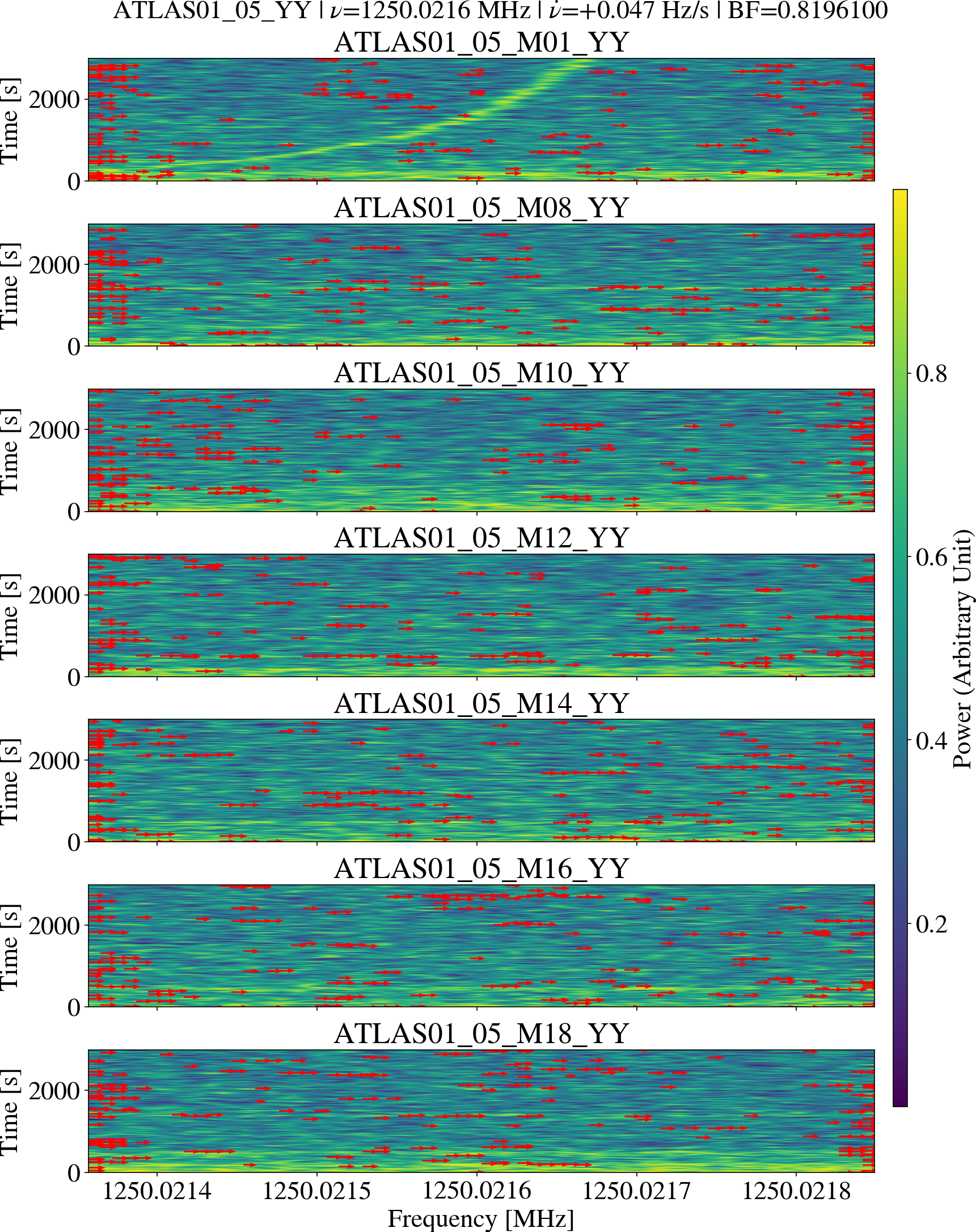}{0.45\textwidth}{(a)}
  \fig{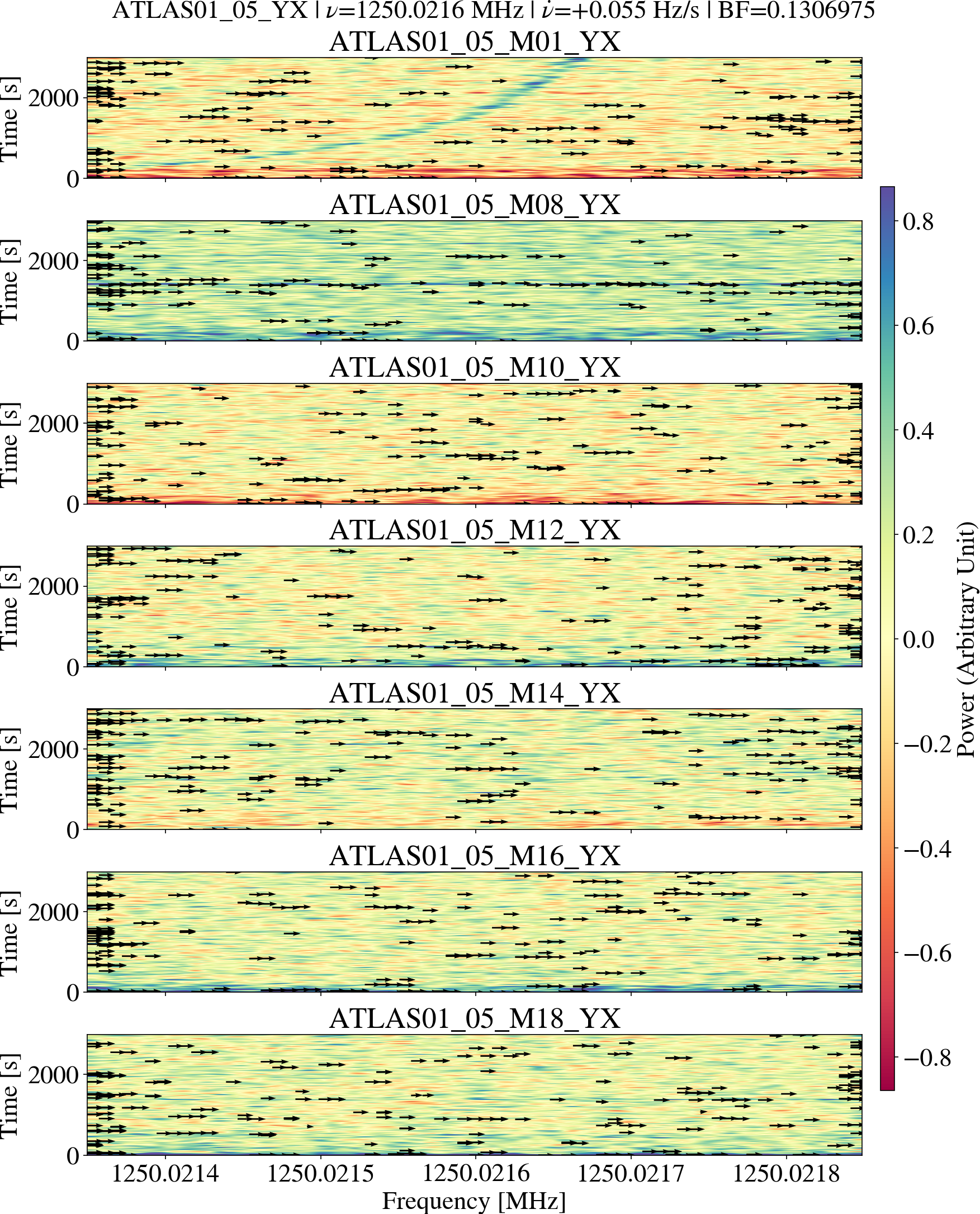}{0.45\textwidth}{(b)}
}\vspace{-6pt}
\gridline{
  \fig{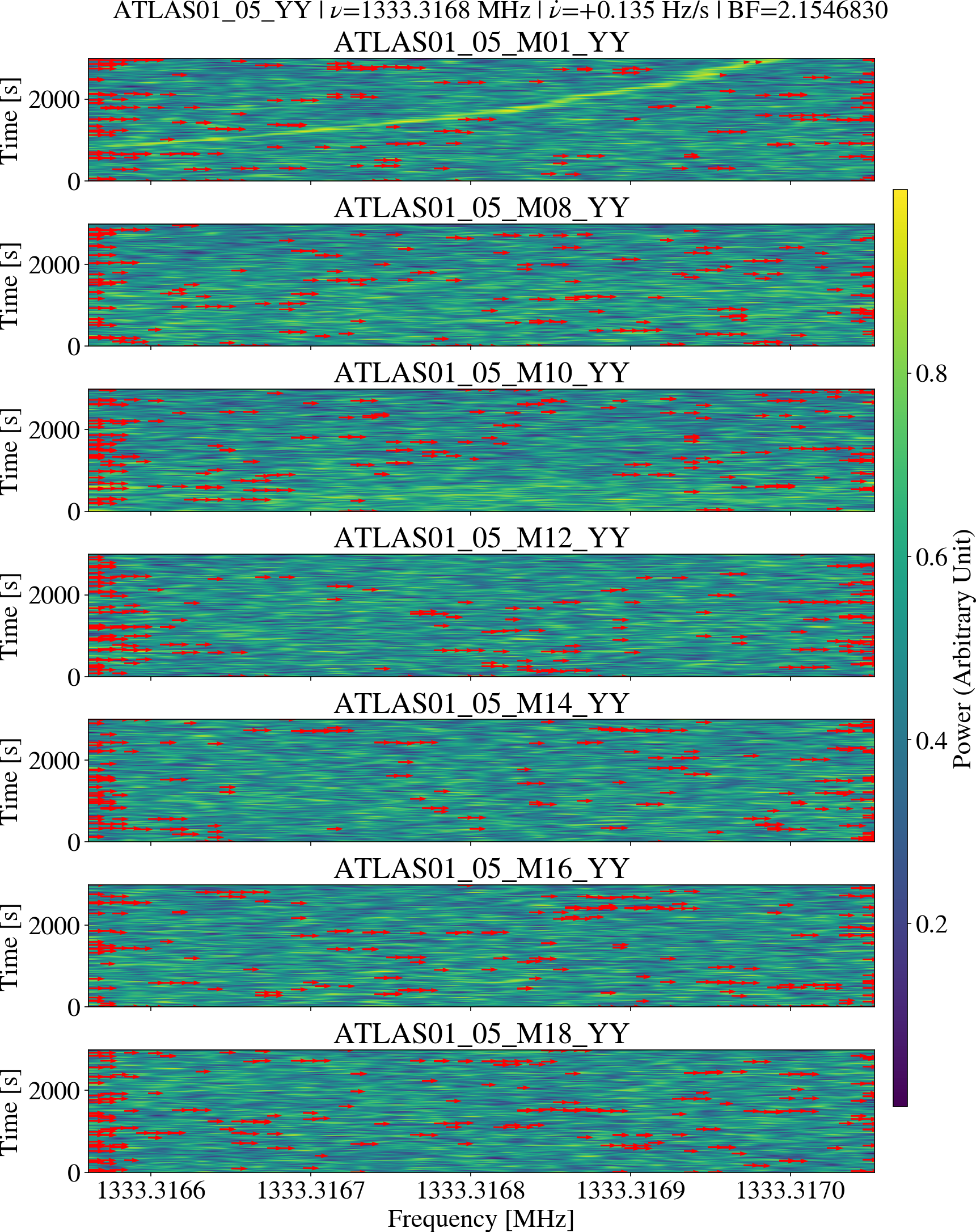}{0.45\textwidth}{(c)}
  \fig{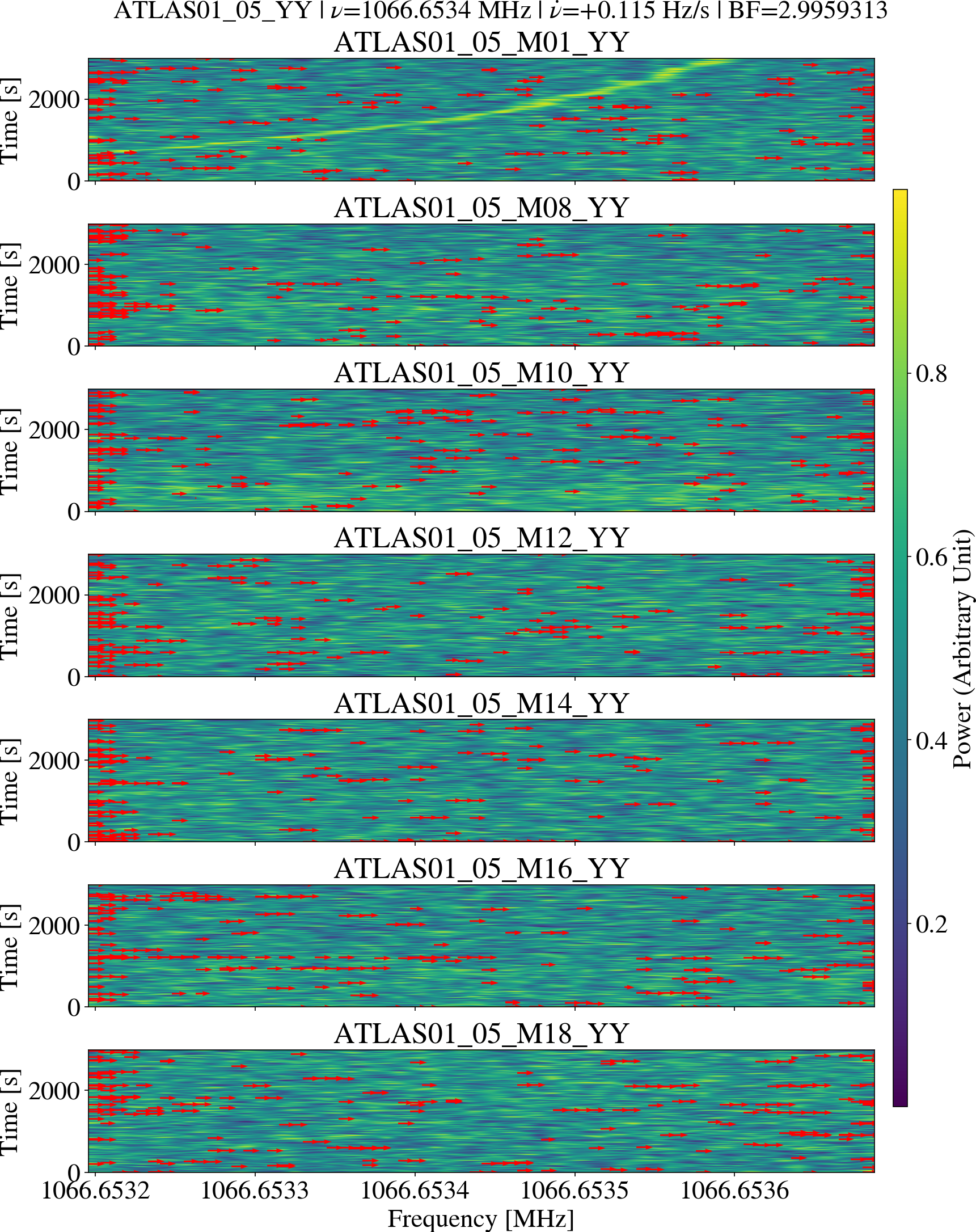}{0.45\textwidth}{(d)}
}
\caption{Four drifting potential candidates with similar chirp-up features with Figure \ref{fig:chirp-up_signal}. Panel (a) and (b) are actually the same signal in YY and YX channels, in spite of different drift rates given by \texttt{bliss}. Panel (c) and (d) are two chirp-up drifting signals with similar chirp-up morphological pattern different from the signals in Panel (a) and (b). Panel (b) uses a differnet diverging colormap to distinguish positive and negative values in the signed cross-hand data.}
\label{fig:four_chirp_up_signal}
\end{figure}

\begin{figure}[htpb]
\centering
\gridline{
  \fig{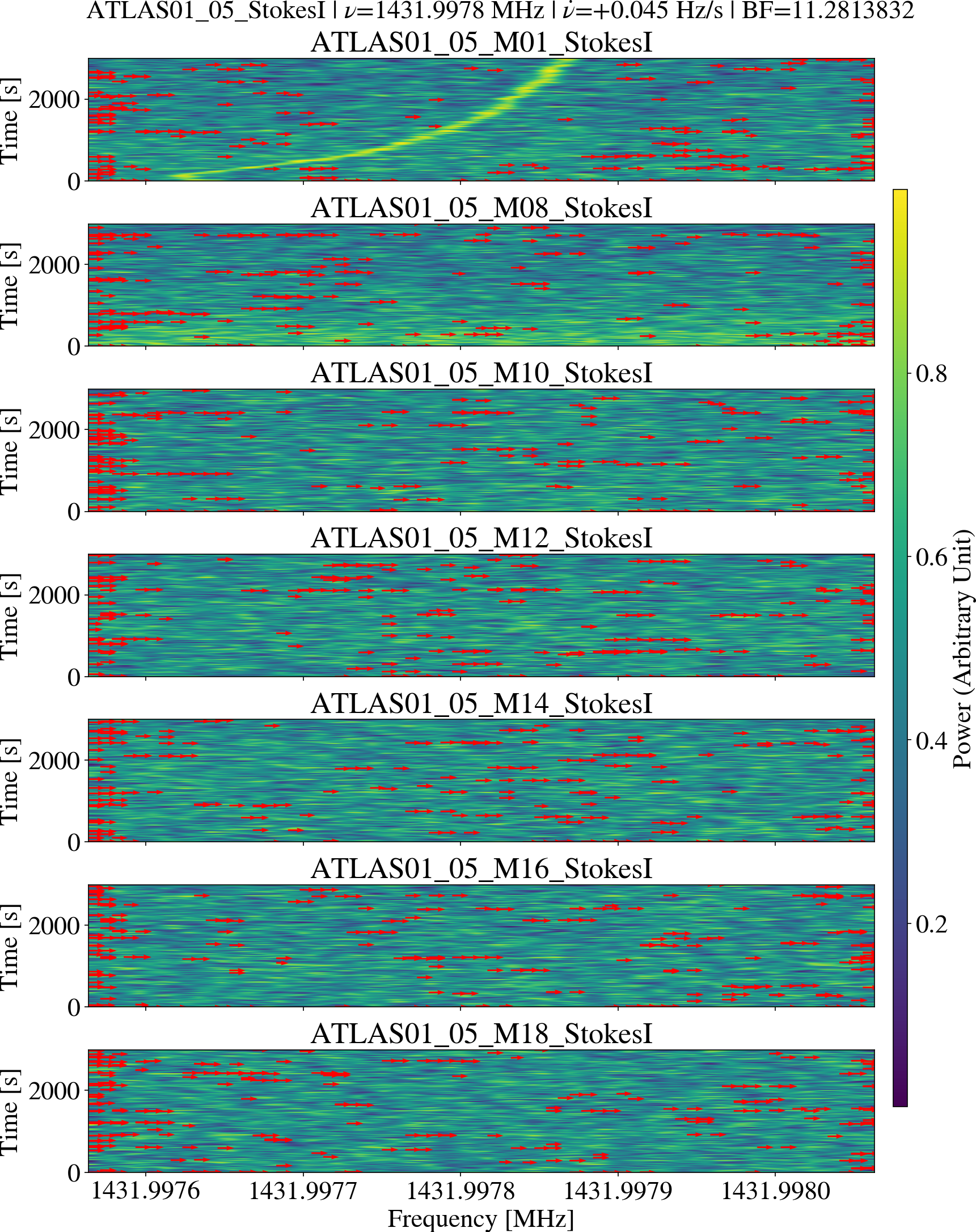}{0.45\textwidth}{(a)}
  \fig{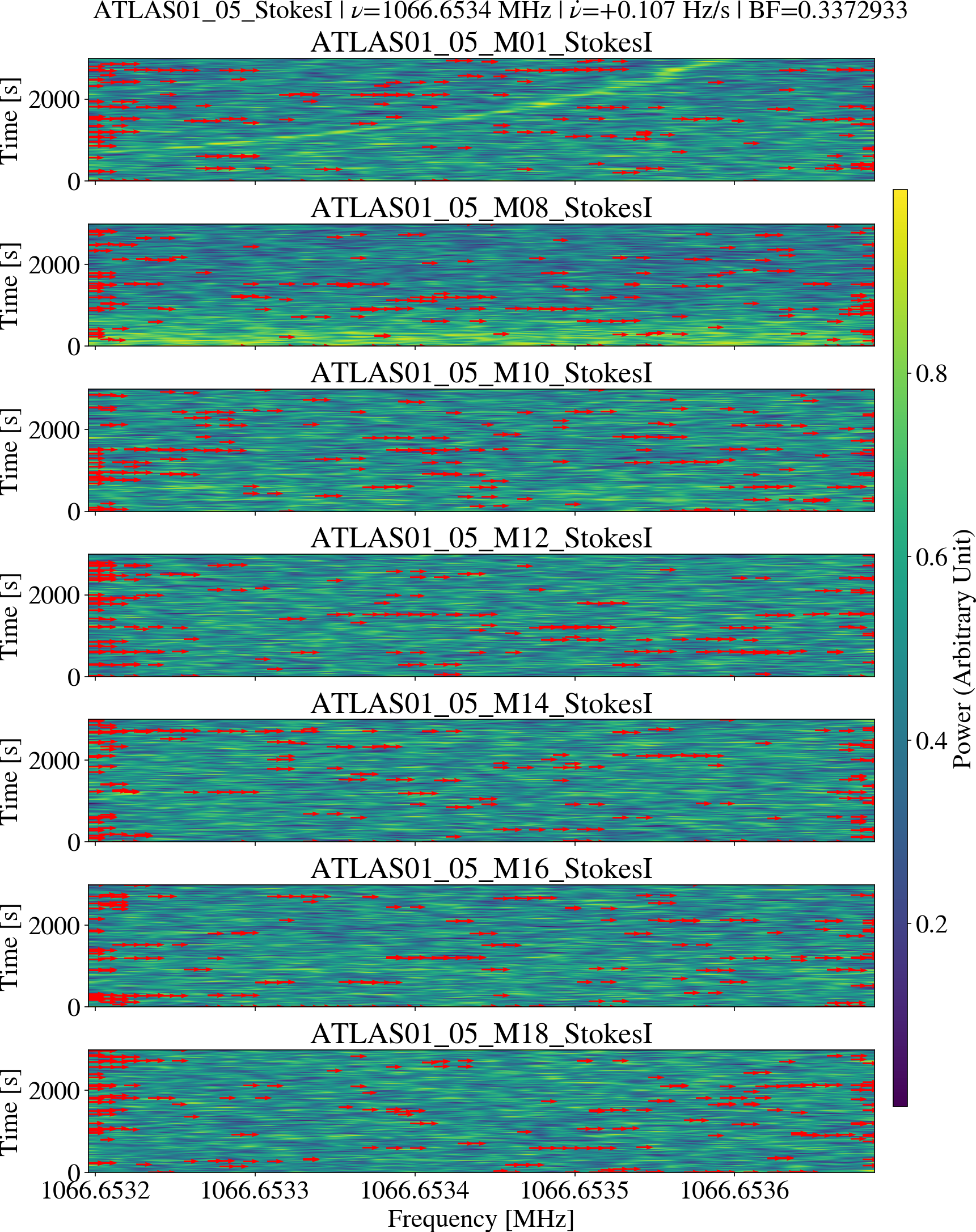}{0.45\textwidth}{(b)}
}
\caption{Two drifting potential candidates with chirp-up features with Figure \ref{fig:chirp-up_signal} in Stokes-$I$. Signal in Panel (a) of this Figure and the signal in Panel (a) of Figure \ref{fig:chirp-up_signal}, as well as signal in Panel (b) of this Figure and the signal in Panel (d) of Figure \ref{fig:four_chirp_up_signal} can be attributed as the same signals dwe to the same frequencies and similar chirp-up shapes.}
\label{fig:two_chirp_up_signal}
\end{figure}

There are seven potential candidates exhibit chirp-up drifting feature. Apart from the potential candidates with event ID 250, other six potential candidates listed in Table \ref{table:cand_list} are illustrated in Figure \ref{fig:four_chirp_up_signal} and Figure \ref{fig:two_chirp_up_signal}.

\section{Derivations for Bayesian Limits}
\subsection{Posterior for Null Detection}\label{subsec:appendix_Posterior}
For the ideal Gaussian noise in radio observation, the test statistic $x$ satisfies $x \sim\mathcal N(\mu_0,\sigma^2)$ for pure noise. For the case of signal detection with EIRP of $L$, the mean of test statistic $x$ becomes $\mu_0+\Delta \mu(L)$, then the SNR can be defined as 
\begin{equation}
  {\rm SNR}=\frac{x-\mu_0}{\sigma_0}.
\end{equation}
The SNR for pure noise satisfies ${\rm SNR}\mid H_0 \sim \mathcal N(0,1)$, while the SNR for signal detection should be ${\rm SNR}\mid H_1 \sim \mathcal N(\Delta \mu(L)/\sigma_0,1)$. The typical SNR for a signal $\mu_{\rm SNR}(L)=\Delta \mu(L)/\sigma_0$ can be given by 
\begin{equation}
  \mu_{\rm SNR}(L)=\frac{\sqrt{\beta}}{{\rm SEFD}}\sqrt{\frac{n_{\rm pol}\tau_{\rm obs}}{\delta \nu_{\rm ch}}}\frac{L}{4\pi d^2}=\zeta L.
\end{equation}
Then the probability $\mathcal{P}_{\rm det}$ of detecting a signal above ${\rm SNR}_{\rm th}$ can be calculated by 
\begin{equation}
  \mathcal P_{\rm det}(L)
=\Pr({\rm SNR}>{\rm SNR}_{\rm th})
=1-\Phi\left({\rm SNR}_{\rm th}-\mu_{\rm SNR}(L)\right)
=\Phi\left(\zeta L-{\rm SNR}_{\rm th}\right).
\end{equation}

We assume $T\mid p_T \sim \mathrm{Bernoulli}(p_T)$, so that $\Pr(T=1\mid p_T)=p_T$ and $\Pr(T=0\mid p_T)=1-p_T$. Conditional on $T=1$, the transmitter has an EIRP of $L$ drawn from a prior $\pi(L)$, while for $T=0$, $L$ is not physically meaningful and will not enter the likelihood below. Consider $K$ observational configurations, where configuration $k$ contains $N_k$ independent trials. For a given EIRP $L$, the single-trial detection probability in configuration $k$ is $\mathcal{P}_{{\rm det},k}(L)$. Using Bayes' theorem and assuming independent priors $\pi(p_T)$ and $\pi(L)$, the joint posterior for $(p_T,L,T)$ is
\begin{equation}
p(p_T,L,T\mid \mathrm{all0})=
\frac{\mathcal{L}(\mathrm{all0}\mid T,L)\pi(L),\Pr(T\mid p_T)\pi(p_T)
}{p(\mathrm{all0})},
\label{eq:joint_post_pTLT}
\end{equation}
where $\Pr(T\mid p_T)=p_T^{T}(1-p_T)^{1-T}$, and the evidence is
\begin{equation}
p(\mathrm{all0})=
\sum_{T\in{0,1}}
\iint\mathcal{L}(\mathrm{all0}\mid T,L)\pi(L)\Pr(T\mid p_T)\pi(p_T)
{\rm d}L{\rm d}p_T.
\label{eq:evidence}
\end{equation}
Finally, we marginalize over the latent indicator $T$ to obtain the posterior in $(p_T,L)$:
\begin{equation}
p(p_T,L\mid \mathrm{all0})=
\sum_{T\in{0,1}} p(p_T,L,T\mid \mathrm{all0})
\frac{\left[(1-p_T)\mathcal{L}(\mathrm{all0}\mid T=0)\right]+\left[p_T\mathcal{L}(\mathrm{all0}\mid T=1,L)\right]}{p(\mathrm{all0})}\pi(L)\pi(p_T).
\label{eq:post_pTL}
\end{equation}
Substituting Equations (\ref{eq:like_all0_given_L})-(\ref{eq:like_all0_T0}) gives a compact mixture-likelihood form:
\begin{equation}
p(p_T,L\mid \mathrm{all0})=
\frac{\left[(1-p_T)+p_T\prod_{k=1}^{K}\left(1-\mathcal{P}_{{\rm det},k}(L)\right)^{N_k}\right]\pi(L)\pi(p_T)}
{\iint{\left[(1-p_T)+p_T\prod_{k=1}^{K}\left(1-\mathcal{P}_{{\rm det},k}(L)\right)^{N_k}\right]\pi(L)\pi(p_T){\rm d}L{\rm d}p_T}}.
\label{eq:post_pTL_final}
\end{equation}

\subsection{Adopted Parameters and Hyperparameters}\label{subsec:appendix_Parameters}
In this Appendix we summarize the prior families and the numerical hyperparameters adopted in the illustrative Bayesian analysis. These choices are intentionally simple and are not intended to represent a unique physical model. Rather, they encode conservative expectations that (i) the transmitter occurrence probability is small but poorly known in scale, and (ii) the EIRP may span orders of magnitude and is therefore more naturally parameterized in $\ln L$.

For the non-hierarchical models, we initially adopt a deliberately loose upper bound $p_{T,\max}=0.5$ to ensure that the prior support covers a broad range of plausible occurrence probabilities and parameterize $p_T=p_{T,\max}q$. We consider (i) a scaled-beta prior $q\sim{\rm Beta}(a,b)$ with $(a,b)=(5,5000)$, which concentrates prior mass near $q\approx a/(a+b)\simeq 10^{-3}$ (thus $p_T\sim 5\times 10^{-4}$) while retaining a long tail to larger values, and logit-normal mapping $p_T=p_{T,\max}\,{\rm sigmoid}(z)$ with $z\sim\mathcal N(\mu_p,\sigma_p^2)$ and $(\mu_p,\sigma_p)=(-3,10)$, which is deliberately diffuse in logit space and therefore weakly informative for $p_T$ over many orders of magnitude. For the EIRP we adopt a log-normal prior $\ln L\sim\mathcal N(\mu_L,\sigma_L^2)$ with $(\mu_L,\sigma_L)=(\ln 2,\ln 10)$, implying a median $L\simeq 2 {\rm W}$ and a multiplicative scatter of $\exp(\sigma_L)=10$.

In the non-hierarchical models, the resulting one-sided upper limits on $p_T$ lie well below the deliberately loose bound $p_{T,\max}=0.5$. Therefore, we subsequently restrict the hierarchical analyses to a grid of smaller $p_{T,\max}$ values, $p_{T,\max}\in\{0.01,0.008,0.005\}$. For the scaled-beta hierarchical model we retain $p_T=p_{T,\max}q$ and use 
\begin{equation}
q\mid m,\kappa \sim {\rm Beta}(m\kappa,\,(1-m)\kappa),\quad
m\sim{\rm Beta}(\alpha_m,\beta_m),\quad
\kappa\sim{\rm Gamma}(a_\kappa,b_\kappa),
\end{equation}
with $(\alpha_m,\beta_m)=(2,2000)$ and $(a_\kappa,b_\kappa)=(5,0.05)$. This places $m \sim 10^{-3}$, pushing the typical $p_T$ to small values, while $\kappa$ controls how tightly $p_T$ is concentrated around $m$: larger $\kappa$ produces stronger shrinkage (narrower posteriors in $\log p_T$) and makes the inferred $(p_T,L)$ region more sensitive to the imposed upper bound $p_{T,\max}$.

For the logit-normal hierarchical model we adopt
\begin{equation}
  z\mid \mu_p,\sigma_p \sim \mathcal N(\mu_p,\sigma_p^2),\quad
p_T=p_{T,\max}\,{\rm sigmoid}(z),
\end{equation}
with $\mu_p\sim\mathcal N(\mu_{p,0},\tau_{\mu_p}^2)$ and $\sigma_p\sim{\rm HalfNormal}(s_{\sigma_p})$. We use $(\mu_{p,0},\tau_{\mu_p})=(-3,4)$ and $s_{\sigma_p}=2$, which yields a broad prior on $p_T$ in logit space but still favors small $p_T$. Compared to the scaled-beta hierarchical model, this mapping tends to redistribute probability more smoothly in $\log p_T$, producing smaller morphological changes of the diagnostic $\log p_T-\log L$ contours when $p_{T,\max}$ is varied.

For the EIRP hierarchical model we retain a log-normal form 
\begin{equation}
  \ln L\mid \mu_L,\sigma_L\sim\mathcal N(\mu_L,\sigma_L^2),\quad  \mu_L\sim\mathcal N(\mu_{L,0},\tau_{\mu_L}^2) ,\quad \ln\sigma_L\sim\mathcal N(\ln\sigma_{L,0},\tau_{\ln\sigma_L}^2). 
\end{equation}
We adopt $(\mu_{L,0},\tau_{\mu_L})=(0,1)$ and $(\ln\sigma_{L,0},\tau_{\ln\sigma_L})=(0.83,0.25)$, corresponding to a prior median $L\sim 1 {\rm W}$ and a typical dispersion $\sigma_L\sim e^{0.83}\approx 2.3$. In the main occurrence probability limits, $L$ and its hyperparameters are treated as nuisance quantities and marginalized over.

\begin{figure}[htpb]
  \centering
  \includegraphics[width=0.85\textwidth]{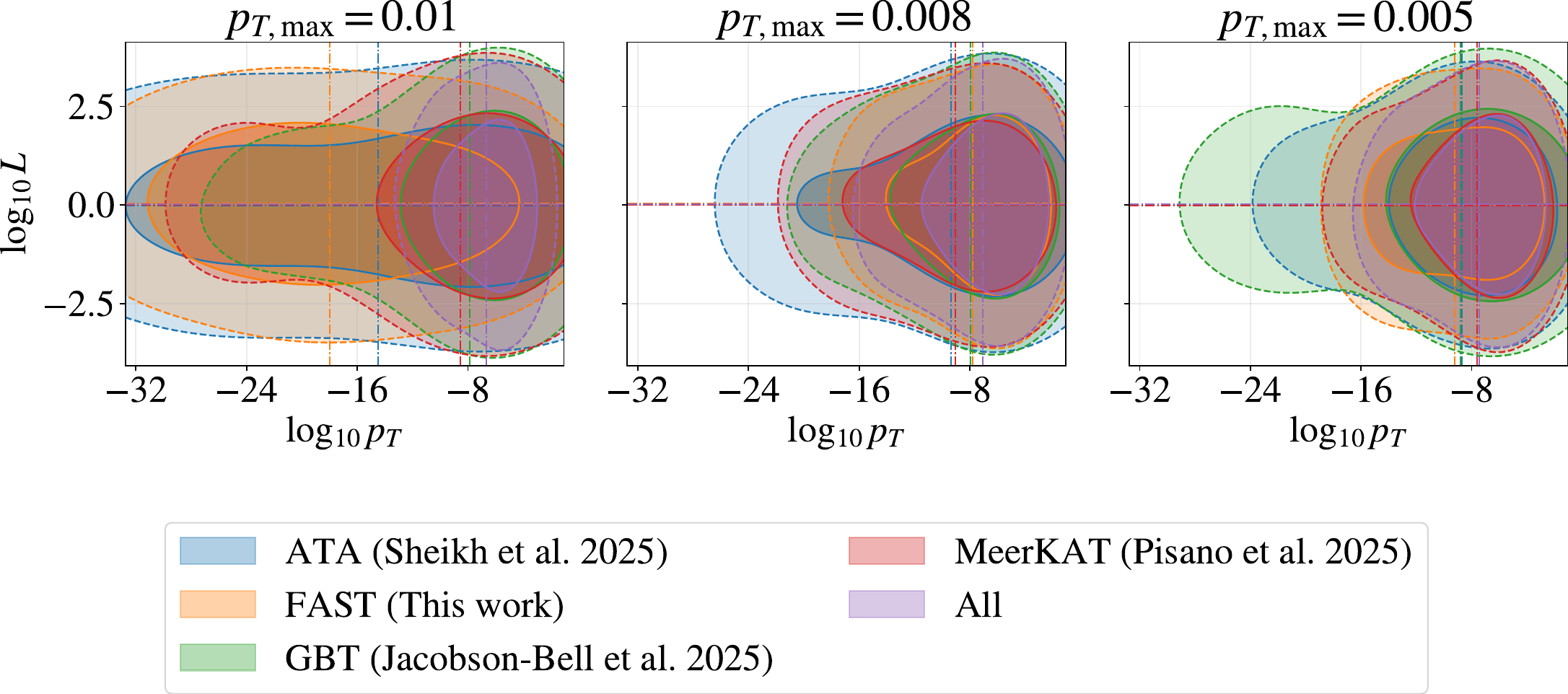}
  \includegraphics[width=0.85\textwidth]{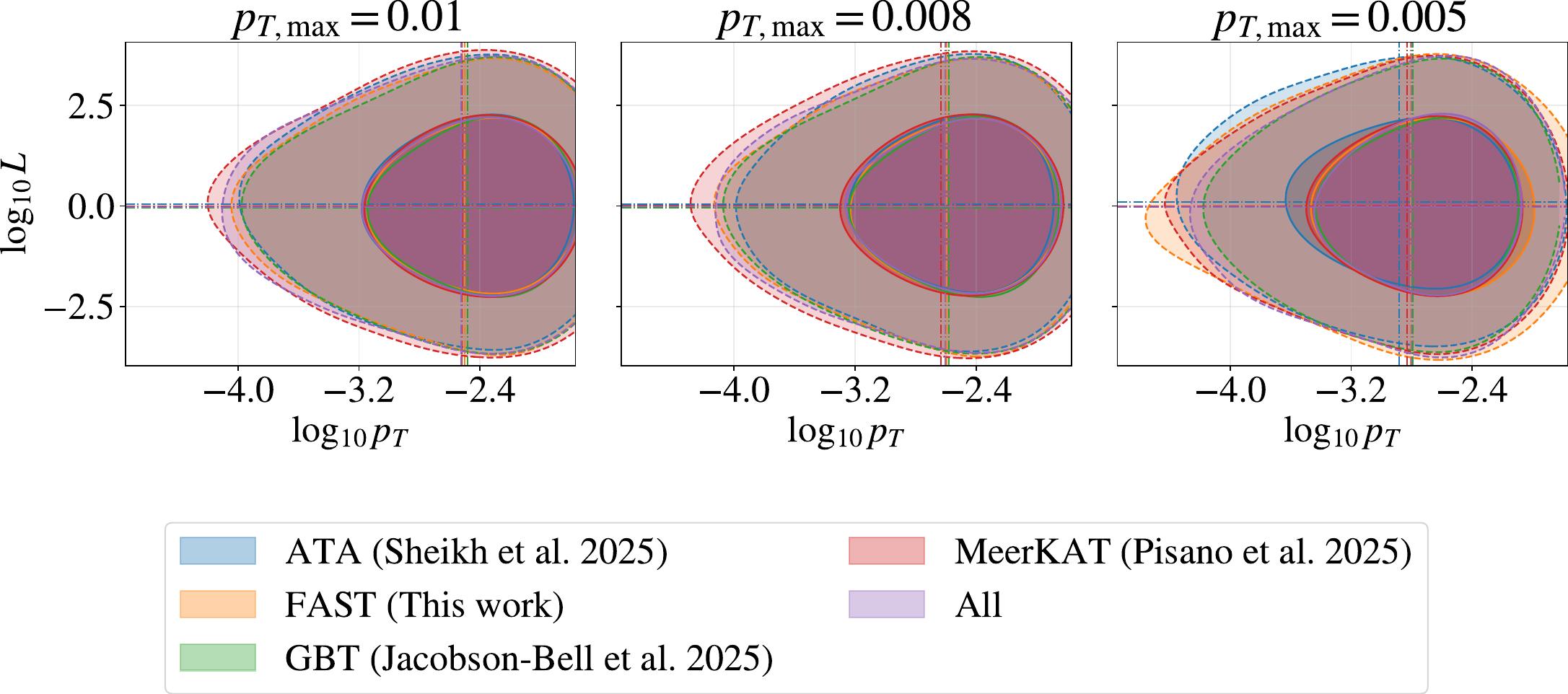}
  \caption{\label{fig:Posterior_logpT_logL}
Diagnostic joint posterior contours for the scaled-beta (top) and scaled logit-normal (bottom) hierarchical prior models with different $p_{T,\max}$ settings. The contours show the joint posterior densities in the $\log_{10}p_T$--$\log_{10}L$ plane after marginalizing over hyperparameters for the observing configurations in ATA, GBT, MeerKAT, and this work. Solid and dashed curves enclose the 68\% and 95\% HPD regions, respectively. }
\end{figure}

\begin{figure}[htpb]
  \centering
  \gridline{
  \fig{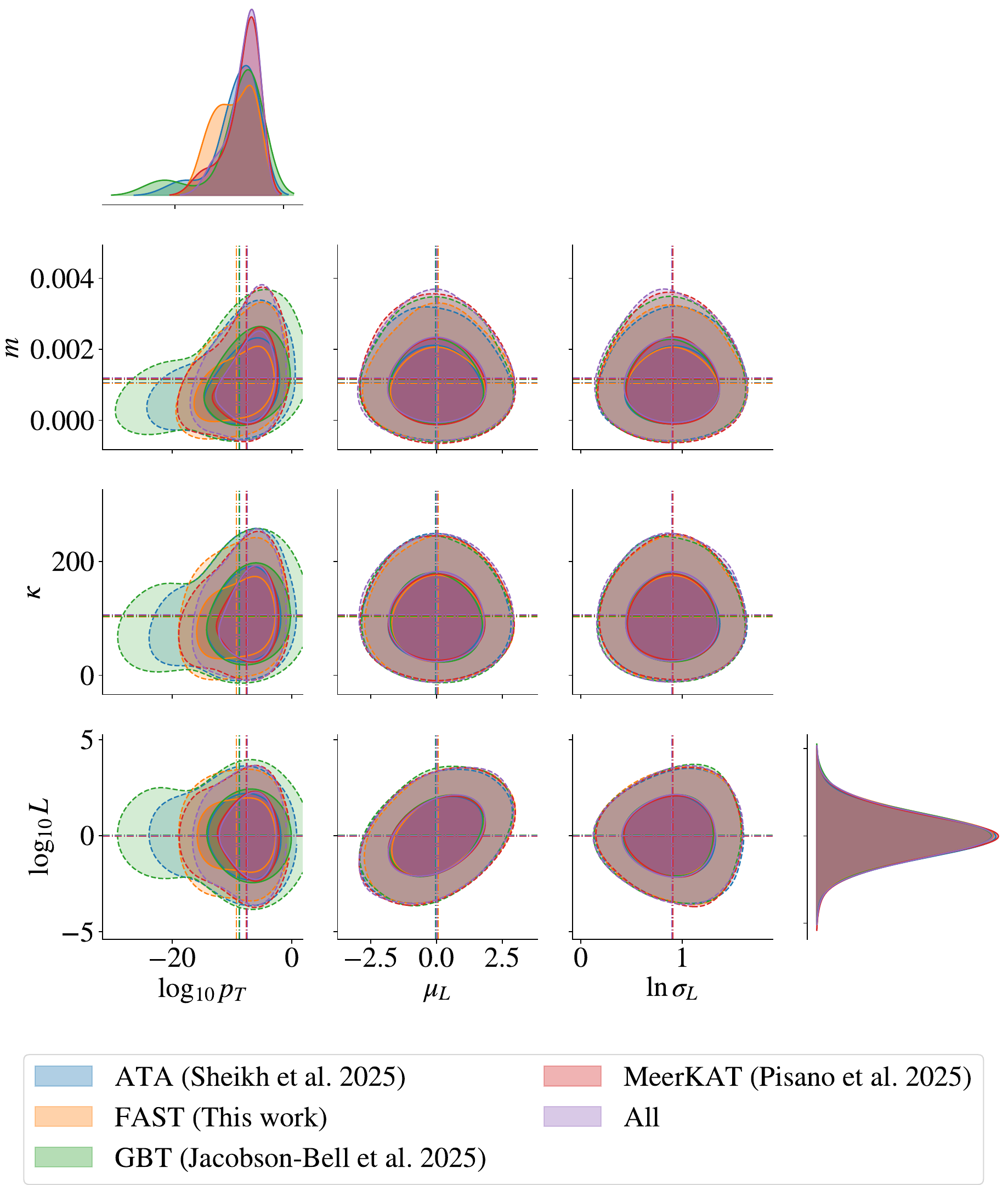}{0.5\textwidth}{(a)}
  \fig{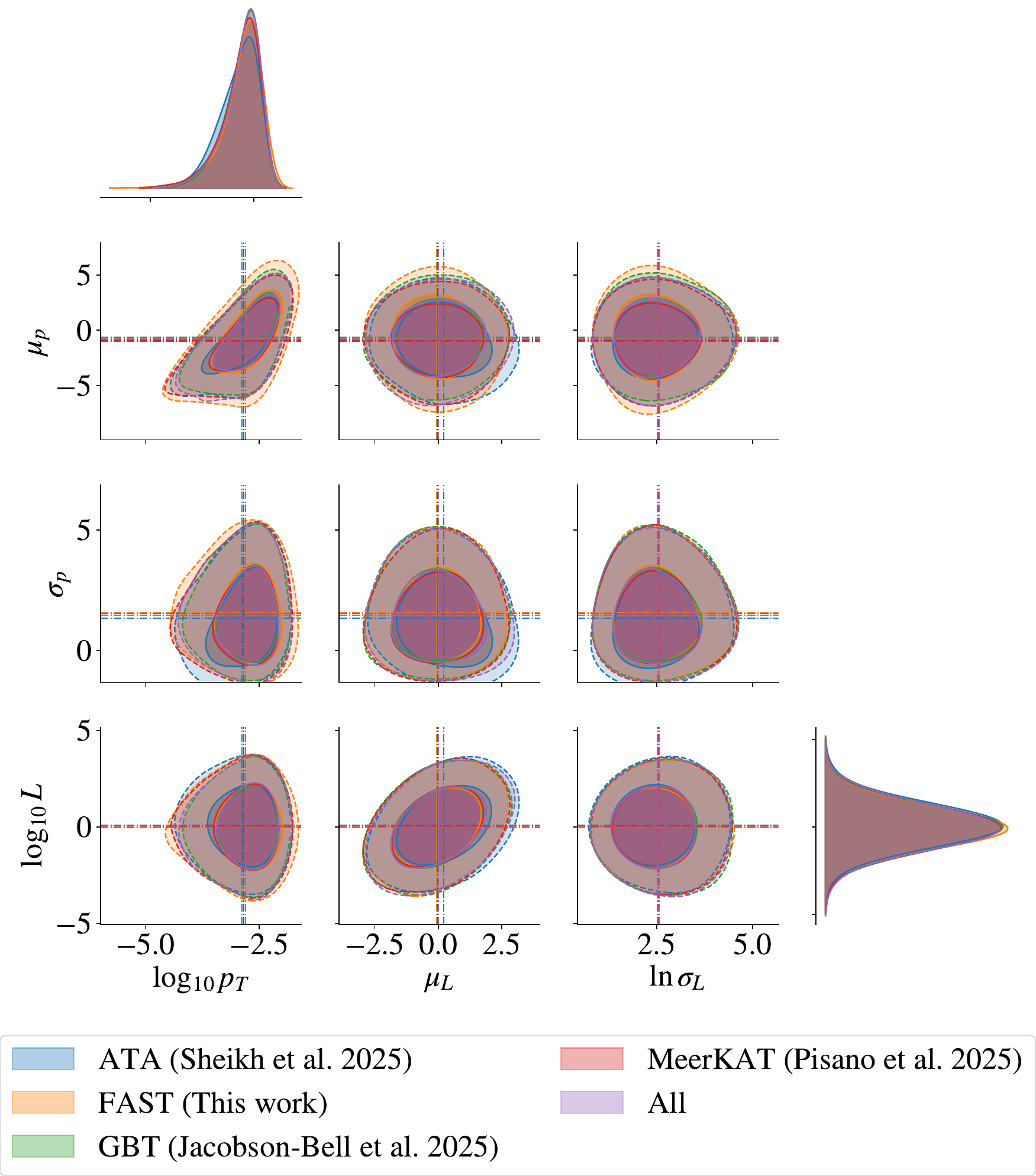}{0.5\textwidth}{(b)}
}
  \caption{\label{fig:Posterior_contour_pT_L}
Diagnostic posterior contours for the scaled-beta (a) and scaled logit-normal (b) hierarchical prior models with $p_{T,\max}=0.005$. The contours show the joint posterior densities in the $\log_{10}p_T$--$\log_{10}L$ plane after marginalizing over hyperparameters for the observing configurations in ATA, GBT, MeerKAT, and this work. Solid and dashed curves enclose the 68\% and 95\% HPD regions, respectively. The top-left panel shows the one-dimensional marginal posterior of $\log_{10}p_T$, while the bottom-right panel shows the one-dimensional marginal posterior of $\log_{10}L$.}
\end{figure}

In the all-zero detection regime, the likelihood provides limited leverage on $p_T$ unless the effective detection probability $f_{\nu,k}\mathcal P_{{\rm det},k}(L)$ varies appreciably across the prior-supported range of $L$. The imposed value of $p_{T,\max}$ therefore affects the allowed range of the marginal posterior of $p_T$, especially for the scaled-beta family. As illustrative examples, Figure~\ref{fig:Posterior_logpT_logL} shows how the diagnostic joint posterior contours change with different choices of $p_{T,\max}$ for the scaled-beta and logit-normal hierarchical models. Figure~\ref{fig:Posterior_contour_pT_L} shows the corresponding diagnostic posterior contours for the adopted case with $p_{T,\max}=0.005$. These figures are included only to visualize the remaining $p_T$--$L$ degeneracy and the prior-family dependence. The numerical upper limits reported in Table~\ref{table:logp_T-logL} are computed from the one-dimensional marginal posterior $p(p_T\mid{\rm all0})$. 

\bibliography{sample631}{}
\bibliographystyle{aasjournal}



\end{document}